\documentclass[journal,transmag]{IEEEtran}
\usepackage[T1]{fontenc}
\usepackage{caption}
\usepackage{cite}
\usepackage[top=2cm, bottom=2cm, left=2cm, right=2cm]{geometry}  
\usepackage{algorithm}  
\usepackage{algorithmicx}  
\usepackage{algpseudocode}
\usepackage{color} 
\usepackage{amsmath}  
\usepackage{setspace}

\usepackage[pdftex]{graphicx}
\usepackage{graphicx}
\usepackage{subfigure}
\usepackage{amsmath}
\usepackage[cmintegrals]{newtxmath}
\usepackage[caption=false,font=normalsize,labelfont=sf,textfont=sf]{subfig}
\usepackage{url}
\begin{document}

\title{A Sequential Variational Mode Decomposition Method}

\author{Wei Chen \\{JiangXi University of Finances and Economics}
\thanks{}
\thanks{}
\thanks{}}

\markboth{}%
{Shell \MakeLowercase{\textit{et al.}}: Bare Demo of IEEEtran.cls for IEEE Communications Society Journals}

\maketitle

\begin{abstract}
To restore the non-stationary signal mixture back to its original components is an interesting topic in the signal processing field. Several methods focusing on this problem have been introduced. "Empirical Mode Decomposition (EMD)" can decompose the mixed signals to some so-called "Intrinsic Mode Functions (IMFs)". However, the EMD method is more like a mathematical trick, lacking in theoretical support. Another algorithm called "Empirical Wavelet Transform (EWT)" has been invented. This method divides the frequency spectrum of the signal into certain pre-defined number of segments empirically, then applies a wavelet transform using some specific wavelet basis. Later, a new method called "Variational Mode Decomposition (VMD)" has been proposed, which is based on the variational method and converts the decomposition issue to an optimization problem. Such an algorithm has a solid mathematical fundamental and shows excellent performance. However, there are also some limitations, mainly including requirement in prior information of component number and end effect.
   
In this paper, we introduce a sequential variational mode decomposition method to separate non-stationary mixed signals successively. This method is inspired by the variational method, and can precisely recover the original components one by one from the raw mixture without prior knowing or assuming the number of components. Such character brings great convenience for real application and differs from the current VMD method. Furthermore, we also conduct a principal elongation for the mixture signal before the decomposing operation. By applying such an approach, the end effect can be reduced to a low level compared with the VMD method. To obtain higher accuracy, a refinement process has been introduced after gross extraction. Combined these techniques together, the final decomposition result implies a significant improvement compared with the VMD method and EMD method.
\end{abstract}


\IEEEpeerreviewmaketitle

\section{Introduction}

\IEEEPARstart{N}{on-stationary} mixed signal decomposition problem is a challenge in signal processing research. Short-time Fourier transform has been applied to process this problem but the window length is difficult to select due to the strong non-stationarity \cite{STFT}. Sometimes, Wavelet transform is used while it is difficult to choose proper wavelet basis functions and such functions must be modified as the input mixed-signal changes, without generality \cite{WaveletTrans_p1}.

Huang et al in 1998 introduced EMD method attempting to solve this problem \cite{EMD}. The algorithm firstly searches the local maximum and minimum points in the mixed-signal and then interpolates these maxima and minima to obtain upper and lower envelopes. After this process, the average of such two envelopes is considered as a narrowband filter and the mixed signal is sifted iteratively until the result satisfies the definition of IMFs. Repeating such operations for the remaining signal until the stop criterion is achieved, a series of IMFs would be extracted. This algorithm indeed supplies a simple but effective technique to separate the signal mixture into principle modes sometimes. Each sifting operation for a single IMF acts like sequential filtering done by a series of adaptive filters whose center frequencies are listed from high to low. However, using EMD, the deviation in the end regions may be very significant. Also if mixed by the noise or intermittence signals, there would be modal aliasing appeared. Besides, this algorithm is basically a mathematical trick, lacking theoretical support \cite{EMD_p1,EMD_p2,EMD_p3}.

Though there are mentioned limitations in the EMD method, it provides a powerful tool in the time-frequency analysis field, especially for non-stationary signal processing. This method has been widely applied in audio engineering, biomedical signal processing, mechanical vibration analysis, etc \cite{EMD_app2,ECG_EMD,EMD_app1}. Some improvements aimed to reduce mode mixing phenomena have been made based on such an algorithm, and related methods are proposed, such as EEMD, CEEMD, FCEEMD, etc \cite{EEMD,CEEMD,FEEMD}. These methods are proposed to eliminate mode aliasing from white noise mainly by adding independent artificial white noise into the original signal repeatedly. The final IMFs are obtained by averaging corresponding IMF components derived from the noise added signals by the EMD method. Obviously, These modified algorithms increase time cost badly.

In 2013, Jerome Gilles proposed the EWT algorithm \cite{EWT}. This method is based on wavelet transform but with selected wavelet basis functions. Given the number of principal modes, the frequency spectrum is divided into corresponding segments centered around certain maxima and with boundaries averaged by consecutive two maxima. Thus the empirical wavelet basis functions are built based on such segmentation. The principal modes are obtained by regular wavelet transform with these basis functions. This approach is adaptive and with strong theoretical support. The decomposition result is comparable with EMD. However, it is still empirical during the construction of wavelet basis and in the number determination of the principal components.

Dragomiretskiy et al submitted a new algorithm named VMD for this problem in 2013 \cite{VMD}. The candidate algorithm is based on the variational method and transfers such decomposition issue to an optimization problem. Hilbert transform would be conducted for the signal before the optimization process. This algorithm characters with solid theoretical fundamental and reasonable assumptions. And from the test, it shows a smaller deviation from the original principal modes and also good robustness. With proper modification, this algorithm can be extended to deal with the decomposition of two-dimensional signals, such as 2D image processing. Though with many advantages, there are also some drawbacks. One is that the number of intrinsic modes must be defined a priori. In fact, such information is not so frequently given. The second one, also existing in EMD and EWT, is the serious end effect. That is to say, utilizing the mentioned three algorithms, there would be a relatively larger deviation between the obtained intrinsic modes and the true components in the regions close to the two ends. The commonly used approach to reduce such effect is mirroring the data close to two ends in VMD, however, such an approach can only maintain the consistency in amplitude but fails in the phase and trend.

In this paper, we proposed a new algorithm named "Sequential Variational Decomposition Method (SVMD)". This method can fetch one component from one decomposition, without prior knowledge or estimation of the number of intrinsic modes. This function is achieved by modifying the optimization form based on VMD. This advantage can widen the application field of the VMD method in non-stationary signal decomposition. This new algorithm performs as a series of adaptive filters to extract principal components from the mixture one by one. To reduce the end effect, an approach called "Principal Component Restoring (PCR)" is introduced. We extract the trend line and several principal components from the end regions, then conduct an elongation operation based on such information for the signal. In the proposed method, we would do decomposition for the elongated signal, instead of the mirrored one in VMD. Also, the extension of its application to two-dimensional signal decomposition can be realized with some essential modifications. 

The following part of this paper would be arranged as follows: Section II firstly would briefly describe preparation for the new proposed algorithm. During this review, a preliminary procedure has been clarified. Then derivation of the new algorithm is introduced, mainly including the building of optimization expression and its recursively solving process. As following, the PCR method is introduced to reduce the end effect. In section III, firstly, the initialization and noise effect would be discussed. 
We mainly study the convergence property under different initialization conditions and noise levels. In the new method, refinement is available, so we also introduce a refining approach and evaluate its improvement. Combined with these associated techniques, the complete decomposition method is formed and applied to analyze various signals, including synthetic signal mixtures and a true biomedical signal. Results of comparing with EMD and VMD would also be shown here. The last section summarizes the complete work reported in this paper, limitations and future possible improvements would be also discussed. The generalization of VMD and SVMD methods also has been addressed to solve the similar types of mode decomposition problems.

\section{Algorithm Description}
\subsection{Narrow bandwidth assumption}
All these decomposition algorithms would first define what is a mode. It is the fundamental issue and determines the final profile of the decomposition components. In early work by Huang, the mode is defined in time domain, oscillates with constraint in the number difference between local extreme and zeros \cite{EMD}. Such modes can represent most of the oscillating signals. However, when signals riding upon trend lines that are monotone and deviate greatly from zero, the recovering result based on such mode assumption is not so acceptable. Then in later work, a more normal definition is given, called "Intrinsic Mode Functions (IMFs)", with mathematical form obtained from modulation theory, that is  
\begin{equation}  
\label{AMFM}
u_k(t)=A_k(t)cos(\phi_k(t))
\end{equation} 
$A_k(t),\phi_k(t)$ are amplitude modulation and phase modulation respectively. Some important additional constraints are also defined, such as instantaneous frequency $\omega_k(t)=\phi_{k}^{'}(t)$ is non-negative, $\omega_k(t)$ and $A_{k}^{'}(t)$ are much smaller than $\phi_k(t)$. 
The later definition gives more detailed restrictions compared with the former one. In fact, all these constraints promise that the IMFs are narrow bandwidth in the frequency domain. Such property also inspires the VMD method and becomes the most important prior assumption during the optimization process \cite{VMD}.  

However, if a trend line(e.g., polynomial of 1 or 2 order) is mixed with other oscillating signals, such form of IMFs cannot describe it. As a consequence, using the EMD method to extract such a trend line would bring great deviation. Since it is frequent for this kind of signal mixture, the definition of IMFs with trigonometric form is limited. We must extend it and find out its basic property.

Considering the possibility of a trend line, we must find a more common property shared by both trigonometric and polynomial form. As mentioned before, the second definition of IMFs promises narrow bandwidth. Coincidentally, in fact, this property also exists in the polynomial trend line. Commonly, the frequency spectrum of polynomials with low order shows a narrow bandwidth, with the peak located at zero. It means that we can give this new definition for an IMF: an intrinsic mode function is a time series with narrow bandwidth in the frequency domain. 

\subsection{Hilbert transform and Fourier transform}
Hilbert transform is a powerful tool in signal processing field \cite{HilbertTrans_App1,HilbertTrans_app2}. During this transform, we conduct convolution operation with the impulse response $h(t)=\frac{1}{\pi t}$ for the signal in time domain,
 \begin{equation} 
 \label{hilbtrans}
 H(f(t))=f(t)*h(t)=\frac{1}{\pi}\int_{-\infty}^{+\infty}\frac{f(v)}{t-v}dv
 \end{equation} 
In our case, the introduction of Hilbert transform is to obtain the analytic form for target real signal. It is defined as 
\begin{equation}
\label{analyfun}
f_A(t)=f(t)+jH(f(t))
\end{equation}
From the definition, we can see that the analytic form of a real signal is constructed by combining its result after Hilbert transform as the imaginary part and the original signal as the real part. This analytic signal has many applications, such as envelope computation and instantaneous signal analysis. However, in this paper, we mainly use its property after the Fourier transform.
\begin{equation}
\label{fouriertrans}
\hat{f}_A(\omega)=F\{H(f(t))\}=\hat{f}(\omega)+j\hat{h}(\omega)\hat{f}(\omega)
\end{equation}
Since $h(t)=\frac{1}{\pi t}$, its Fourier transform is $-jsign(\omega)$, so the Equation \ref{fouriertrans} can be rewritten into
\begin{equation}
\label{hilfourier}
\hat{f}_A(\omega)=
\begin{cases}
2\hat{f}(\omega),   &\omega>0\\
0,                  &\omega<0
\end{cases}
\end{equation}
It means that if we want to conduct some operations to the frequency spectrum of the signal after the Hilbert transform and Fourier transform, only the positive frequency region is needed to concern, where the amplitude value is double of that after only Fourier transform.

To decompose the mixed signals, in our method, the original signal of the time domain would be firstly transferred to the frequency domain, and then back to the time domain. The decomposition operation happens in the frequency domain, we must do "Inverse Fourier Transform (IFT)" for the obtained components. So, due to the analytic form, we will just extract the real part after the IFT, that is
\begin{equation}
\label{ifftrans}
f_r(t)=\mathcal{R}e\{IFFT\{\hat{f}_{A}(\omega)\}\}
\end{equation}
$f_r(t)$ is the recovered signal in time domain.
\subsection{Proposed New Algorithm}
After these preparations, we proposed a totally intrinsic, self-adaptive, and general variational method. This algorithm is based on narrow bandwidth assumption, decomposes the mixed signal into so-called IMFs by an optimizing process. Compared with current related approaches, this novel algorithm characters with these breakthroughs: 1) With solid mathematical base compared with EMD; 2) Without prior assumption in the number of principal modes; 3) Improvement in reducing end effect; 4) Sequential decomposition and flexible stop criterion; 5) Refinement possibility, thus higher accuracy; 6) High resistance to noise; 7) Manipulatable output order. Though from the final result, the EMD method can be equalized with an adaptive filter bank \cite{EMD_p3}, it still lacks a theoretical explanation. While the EWT method requires more prior information, such as band boundaries and numbers \cite{EWT_p1}, and it also defines many parameters empirically during the construction of wavelet basis functions \cite{EWT}. The VMD algorithm, which is the startup of our method, must define firstly the number of principal modes, while this information is not so common to be given. And sometimes, not all the principal modes are mandatory, but with a minimum power threshold or specific frequency limitation. This algorithm starts from the narrow bandwidth assumption, utilizes Hilbert transform and Fourier transform to convert the signal into the frequency domain, then conducts a decomposition process based on the optimization. What differs the most from the VMD method is the intrinsic modes are extracted sequentially, that is one mode in one-time decomposition, following a user-defined order. Naturally, the optimization formula also should be modified. 

To realize the sequential extraction, two optimization terms can be derived, that is power term and bandwidth term. The first term could be written as,
\begin{equation}
\label{firstterm}
\left\|\hat{f}_{A,i-1}(\omega)-\hat{u}_{A,i}(\omega)\right\|_2^2
\end{equation}
$\hat{f}_{A,i-1}(\omega)$ is the remaining signal after $(i-1)^{th}$ decomposition. If $i=1$, it means  the original signal. $\hat{u}_{A,i}(\omega)\ $ is the $i^{th}$ component to be extracted.
Also the narrow bandwidth requirement can be expressed as,
\begin{equation}
\label{secondterm}
\left\|\hat{u}_{A,i}(\omega)(\omega-{\omega}_{c,i})\right\|_2^2
\end{equation} 
with the center frequency ${\omega}_{c,i}$ of the component $\hat{u}_{A,i}(\omega)$ equal to
\begin{equation}
\label{centercom}
\omega_{c,i}=\frac{\int_{0}^{+\infty}{\omega}|\hat{u}_{A,i}(\omega)|^2 d\omega}{\int_{0}^{+\infty}|\hat{u}_{A,i}(\omega)|^2 d\omega}
\end{equation}
The bandwidth can be expressed as the sum of distance away from the center frequency with the power amplitude as the weighting factors in the whole spectrum.

It seems the listed two terms are enough to achieve the purpose of extracting the target principal component. However, till now, only the current target principal mode has been considered, the remaining modes are neglected. We must also promise the narrow bandwidth property for the remaining modes. So, the third term is proposed as
\begin{equation}
\label{thirdterm}
\left\|\hat{f}^{r}_{A,i}(\omega)(\omega-{\omega}^{r}_{c,i})\right\|_2^2
\end{equation}
In this equation, $\hat{f}^{r}_{A,i}(\omega)$ is the remaining signal, can be expressed as
\begin{equation}
\label{remaining}
\hat{f}^{r}_{A,i}(\omega)=\hat{f}^{r}_{A,i-1}(\omega)-\hat{u}_{A,i}(\omega)=\hat{f}_{A}(\omega)-\sum_{j=1}^{i-1}\hat{u}_{A,j}(\omega)
\end{equation}
Also ${\omega}^{r}_{c,i}$ is the center frequency of $\hat{f}^{r}_{A,i}(\omega)$, similarly can be shown as
\begin{equation}
\label{centerrest}
{\omega}^{r}_{c,i}=\frac{\int_{0}^{+\infty}{\omega}|\hat{f}^{r}_{A,i}(\omega)|^2 d\omega}{\int_{0}^{+\infty}|\hat{f}^{r}_{A,i}(\omega)|^2 d\omega}
\end{equation}

In fact, the idea of introducing the third term is inspired by the dichotomy. Each time of decomposition, we can consider separating the original signal into two principal modes, the bandwidth of both modes must be as narrow as possible. So the introduction of the third term is reasonable and essential. It is an interesting topic to generalize such method to deal with other similar problems. In fact, we have obtained some relevant encouraging achievements, which will be exhibited in future literature. In this paper, we would just show its effectiveness in decomposing non-stationary mixed signals. As a consequence, the final target equation for the optimization can be written into,
\begin{equation}
\label{totalterm}
\begin{split}
L(\hat{u}_{A,i})=\left\|\hat{f}^{r}_{A,i-1}(\omega)-\hat{u}_{A,i}(\omega)\right\|_2^2+\alpha\left\|\hat{u}_{A,i}(\omega)(\omega-{\omega}_{c,i})\right\|_2^2 \\
+\beta\left\|\hat{f}^{r}_{A,i}(\omega)(\omega-{\omega}^{r}_{c,i})\right\|_2^2
\end{split}
\end{equation}
In this equation, $\alpha$ and $\beta$ are penalty factors. In our case, their ratio ${\beta}/{\alpha} $ would actually affect the result. So we always choose $\alpha=1$, leave only $\beta$ tunable. Our purpose is to minimize $L(\hat{u}_{A,i})$, which is a simple convex optimization problem. Replace   $\hat{f}^{r}_{A,i}(\omega)$ with $\hat{f}^{r}_{A,i-1}(\omega)-\hat{u}_{A,i}(\omega)$ to emerge $\hat{u}_{A,i}(\omega)$ term, then do the differential about $\hat{u}_{A,i}(\omega)$ to $L(\hat{u}_{A,i})$ and set the result to zero, we obtain the solution,
\begin{equation}
\label{solutionC}
\hat{u}_{A,i}(\omega)=\frac{\hat{f}^{r}_{A,i-1}(\omega)(1+\beta(\omega-{\omega}^{r}_{c,i})^2}{1+\alpha(\omega-{\omega}_{c,i})^2+\beta(\omega-{\omega}^{r}_{c,i})^2}
\end{equation}
In this formula, $\hat{f}^{r}_{A,i-1}(\omega)$, ${\omega}^{r}_{c,i}$ and ${\omega}_{c,i}$ are defined in \eqref{remaining}, \eqref{centerrest}, \eqref{centercom} respectively.

Since ${\omega}_{c,i}$ and ${\omega}^{r}_{c,i}$ are both related to $\hat{u}_{A,i}(\omega)$, we can iterate to obtain a solution starting from an initial value noted as $\hat{u}_{A,i}^{0}(\omega)$. The complete procedure is described in Algorithm 1. The outer and inner stop threshold $\epsilon$ and  $\eta$ are defined by user.
\floatname{algorithm}{Algorithm} 
	\begin{algorithm}  	
		\caption{Optimization procedure in SVMD }  
		\begin{algorithmic}[1]  
		\State Initialize $\hat{f}^{r}_{A,0}(\omega)=\hat{f}_{A}(\omega)$ and $\epsilon$;
		\State $i=1$;
		\While {$\left\|\hat{f}^{r}_{A,i}(\omega)\right\|_{2}^2>\epsilon$}  
	     \State Initialize $\hat{u}_{A,i}^{0}(\omega)$ and $\eta$ ;
	     \State $k=1$;
	     \While{$\left\|\hat{u}^{k}_{A,i}(\omega)-\hat{u}^{k-1}_{A,i}(\omega)\right\|_{2}^2>\eta$} 
	     \State $\omega_{c,i}^{k-1}=\frac{\int_{0}^{+\infty}{\omega}|\hat{u}^{k-1}_{A,i}(\omega)|^2 d\omega}{\int_{0}^{+\infty}|\hat{u}^{k-1}_{A,i}(\omega)|^2 d\omega}$;
	     \State $\hat{f}^{r,k-1}_{A,i}(\omega)=\hat{f}^{r}_{A,i-1}(\omega)-\hat{u}^{k-1}_{A,i}(\omega)$;
	     \State $\omega^{r,k-1}_{c,i}=\frac{\int_{0}^{+\infty}{\omega}|\hat{f}^{r,k-1}_{A,i}(\omega)|^2 d\omega}{\int_{0}^{+\infty}|\hat{f}^{r,k-1}_{A,i}(\omega)|^2 d\omega}$;
	     \State $\hat{u}^{k}_{A,i}(\omega)=\frac{\hat{f}^{r}_{A,i-1}(\omega)(1+\beta(\omega-{\omega}^{r,k-1}_{c,i})^2}{1+\alpha(\omega-{\omega}^{k-1}_{c,i})^2+\beta(\omega-{\omega}^{r,k-1}_{c,i})^2}$;
	     \State $k=k+1$;
	     \EndWhile
	     \State $\hat{u}_{A,i}(\omega)=\hat{u}^{k}_{A,i}(\omega)$;
	     \State $\hat{f}^{r}_{A,i}(\omega)=\hat{f}^{r}_{A,i-1}(\omega)-\hat{u}_{A,i}(\omega)$;
	     \State $i=i+1$;
	     \EndWhile      
		\end{algorithmic}  
	\end{algorithm}

From the description in Algorithm 1, we can see that the principal modes are extracted sequentially, whose number is not mandatory being informed beforehand. The algorithm is completely self-adaptive compared with VMD. In fact, when confronting with an unknown mixed signal, the number of modes can not be easily determined. If using VMD, the simplest solution to obtain this parameter is a trial and error way. Obviously, it costs time. There are also other methods proposed to determine the mode number in advance, such as MOPSO, DFA, MCFO, CC, MMI and CFSA method \cite{VMD_p1,VMD_p2,VMD_p3}, but all of them have their limitations. Furthermore, in our method, the mode output order can be manipulated as the user's requirement. In fact, its order is determined by the initialization of $\hat{u}_{A}$. Though in this paper, we follow the amplitude descending order. To realize this purpose, we can set the highest peak of the mixture spectrum as the initial value of $\hat{u}_{A}$ each time. This property also is very meaningful. Since sometimes, the interest for each mode differs, the trend line and components with a higher power are more frequently concerned. Utilizing this algorithm can reduce the time spending on unwanted components. In contrast, when with noise interference, in EMD, the signal that comes firstly from the mixture is always the high-frequency noise, which is useless in most cases. Since the outer stop threshold $\epsilon$, which determines the final number of components, can be modified by the user, it provides the possibility of refinement for the components. The statement contains two senses, one is that lowering $\epsilon$ means more components would be extracted and thus more details of the mixture would emerge; the other is that among the obtained components, we can merge these with close center frequency, which can compensate for the loss of main components extracted in earlier cycle. The relevant experimental results would be shown in the later section.

\subsection{End Effect Reduction}
From the observation, both VMD and EMD algorithms will bring serious deviation between the true signal and extracted components in the two end regions. Generally, the deviation would decrease as approaching the middle. Since the original signal is always non-periodic and finite-length, in VMD, recovering after the Hilbert transform and Fourier transform, there would be distortion in the two ends. And in the EMD algorithm, the rule of extreme judging for ending points is compromised, because signals in the two ends are unilateral, such that the obtained extreme points at the ends maybe not correct. Summarily, deviations in the two algorithms can not be neglected, such "pollution" induced by the end effect would spread to the center region and may accumulate gradually as the decomposing process.

To suppress this phenomenon, several techniques are proposed \cite{EMD_endeffect1,EMD_endeffect2}. These methods all tend to predict the future or past value of the original signal following some specific rules, though achieve some progress, but always in cost of time. Meanwhile, in the original VMD algorithm, the signal is simply mirrored at two ends to keep continuity after elongation, which is far from enough. In fact, to keep coherent with the original signal, the elongated parts must share the same trend with the original one, maintain the same amplitude at the interface points, and oscillate at similar frequencies. 

To maintain the coherence, we proposed a new elongation approach noted as PCR to elongate the signal. This method considers both the time domain and frequency domain. In the time domain, a local trend would be estimated for the short-time end regions using the conventional method (of course, if a whole-region trend is significant, a global trend evaluation also works.). Then, deducing this trend component, we will select several principal components for the remaining signal from the frequency domain. Subsequently, combined the trend line with the principal components, elongation for these two end regions can be realized.

This elongation method balances time-cost and precision. From the tests in the later section, it helps our algorithm show an excellent performance, especially when the original signal rides on a simple polynomial trend line.

\begin{figure}
	\centering
	\includegraphics[width=2.5in]{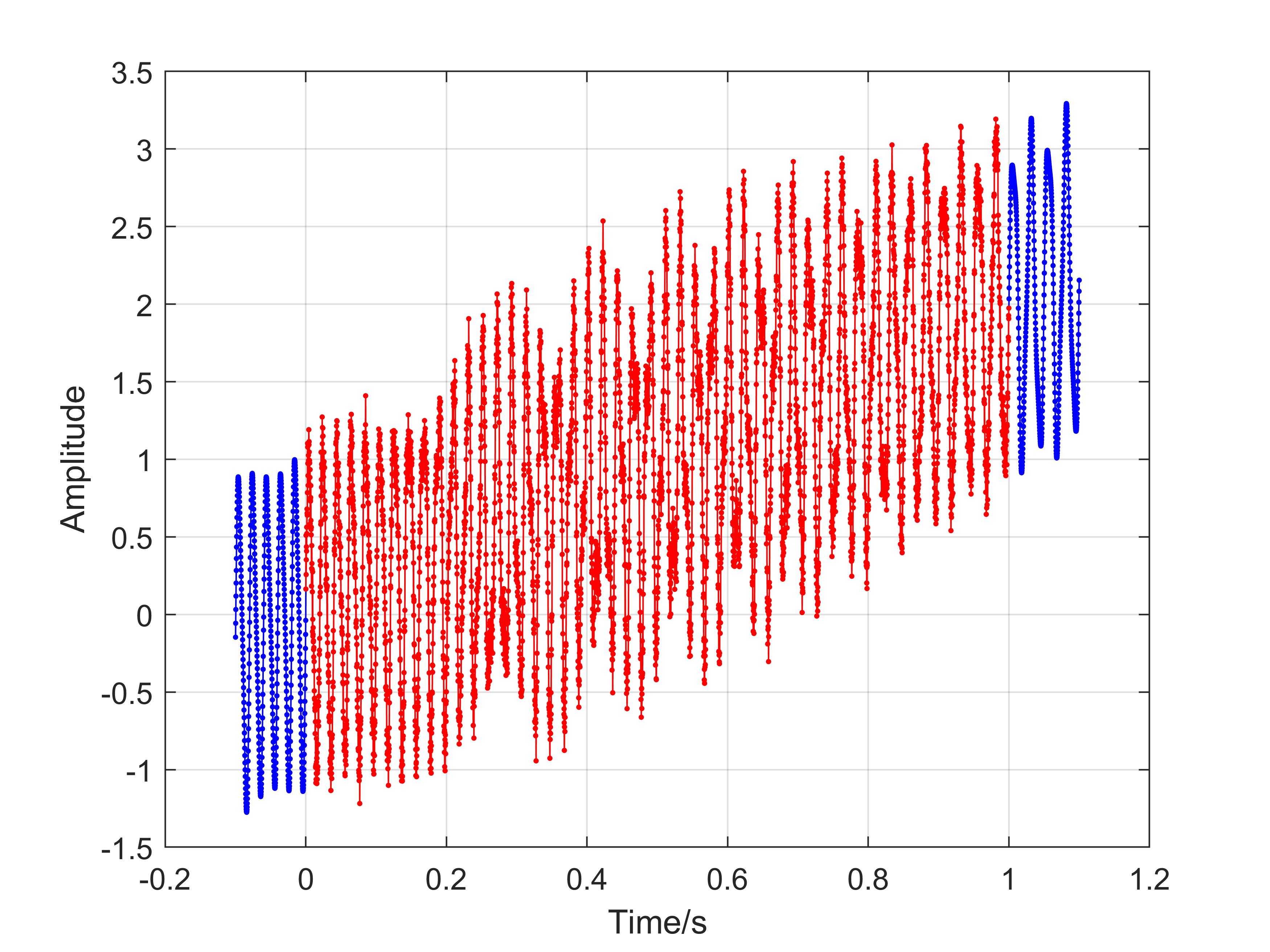}	
	\caption{Elongation operation for original signal with PCR method}
	\label{elongation}
\end{figure}

Fig. \ref{elongation} shows an example of the PCR method used in the elongation for the original signal(red), which rides on a linear trend line. From the figure, we can see that the elongated signal(blue) not only maintains continuity in two end interfaces but also shares the same trend with the original signal.

\section{Experiments and Results}
During this section, we will mainly apply our novel algorithm to decompose a variety of mixed-mode signals. Its performances and properties would be exhibited and examined in detail. Firstly, the initialization problem will be discussed. It affects convergence both in speed and accuracy for the decomposing. Additionally, the noise level is also an important factor affecting decomposition performance. Thus noise susceptibility of the proposed algorithm in convergent range and final limit deviation would be investigated here, relevant comparison with VMD will also be shown. Refinement can help improve decomposition quality. The corresponding experiment also will be conducted by tuning down the stop threshold and then adding components with similar center frequency together. In our approach, it is unnecessary in prior knowledge about the number of modes and the complete series of modes are supposed to be extracted in sequence. To verify this capability, we apply this algorithm to different types of non-stationary signals with various numbers of modes, and compare the results with the VMD method in accuracy and end effect.

For convenience to compare, we limit signals in the time interval from 0 to 1s with a sampling rate 5kHz in all cases. Since we focus on the non-stationary signal processing, the examples listed exclude the pure harmonic signals, thus the tone separating capability will not be studied here.

\subsection{Initialization and noise effect on convergence}
In our case, deduced from the optimization function in \eqref{totalterm}, modes are supposed to be extracted sequentially. However, to achieve this target, a proper set-up of initialization condition is mandatory, since our algorithm cannot guarantee a global optimum. During this section, firstly, we will show how the initial condition affects the convergence and accuracy by studying a concrete example.

\begin{figure}
	\centering
	\includegraphics[width=2.5in]{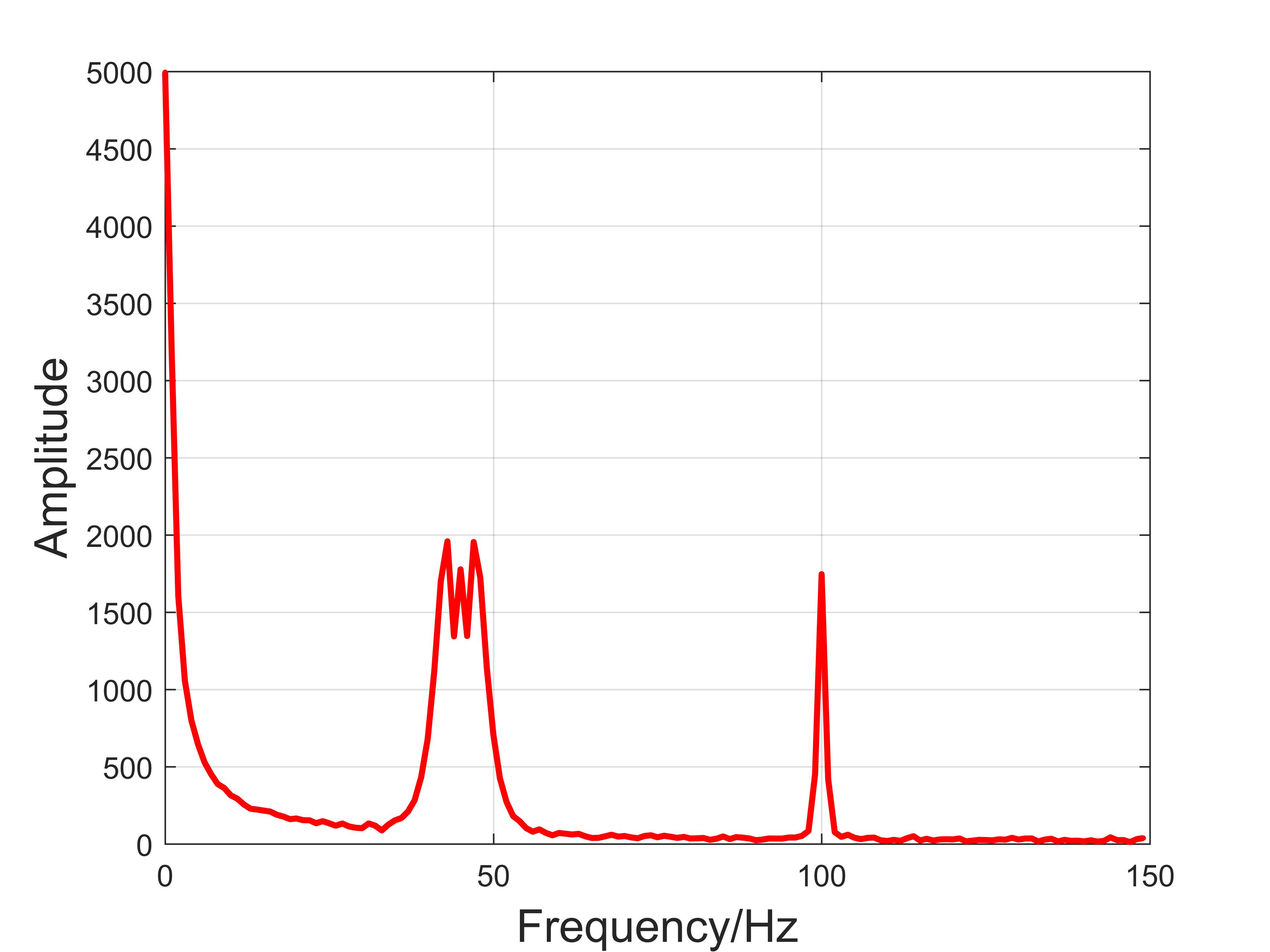}
	\caption{Frequency spectrum of signal 1}
	\label{Spectrum1}
\end{figure}

To keep consistency, we still use signal 1 in section II, whose waveform in the time domain is shown in Fig. \ref{elongation}, and its mathematical expression is
\begin{equation}
f(t)=2t+{\rm sin}(100\pi t-10\pi t^2)+0.5e^{-5(t-0.5)^2}{\rm sin}(200\pi t)+\varepsilon;
\label{sig1}
\end{equation}
Where $\varepsilon$ is the white Gaussian noise with zero mean and variance $\sigma^2$. In the current case, we firstly choose $\sigma$=0.1. Since the optimizing is in the frequency domain, also its spectrum is plotted in Fig. \ref{Spectrum1}. The initializing approach used here is to use a series of impulse train, whose peak positions are set close to the peaks, and corresponding amplitudes are random values.

\begin{figure}
	\centering
	\subfigure[Convergence for C1]{\includegraphics[width=3.5in]{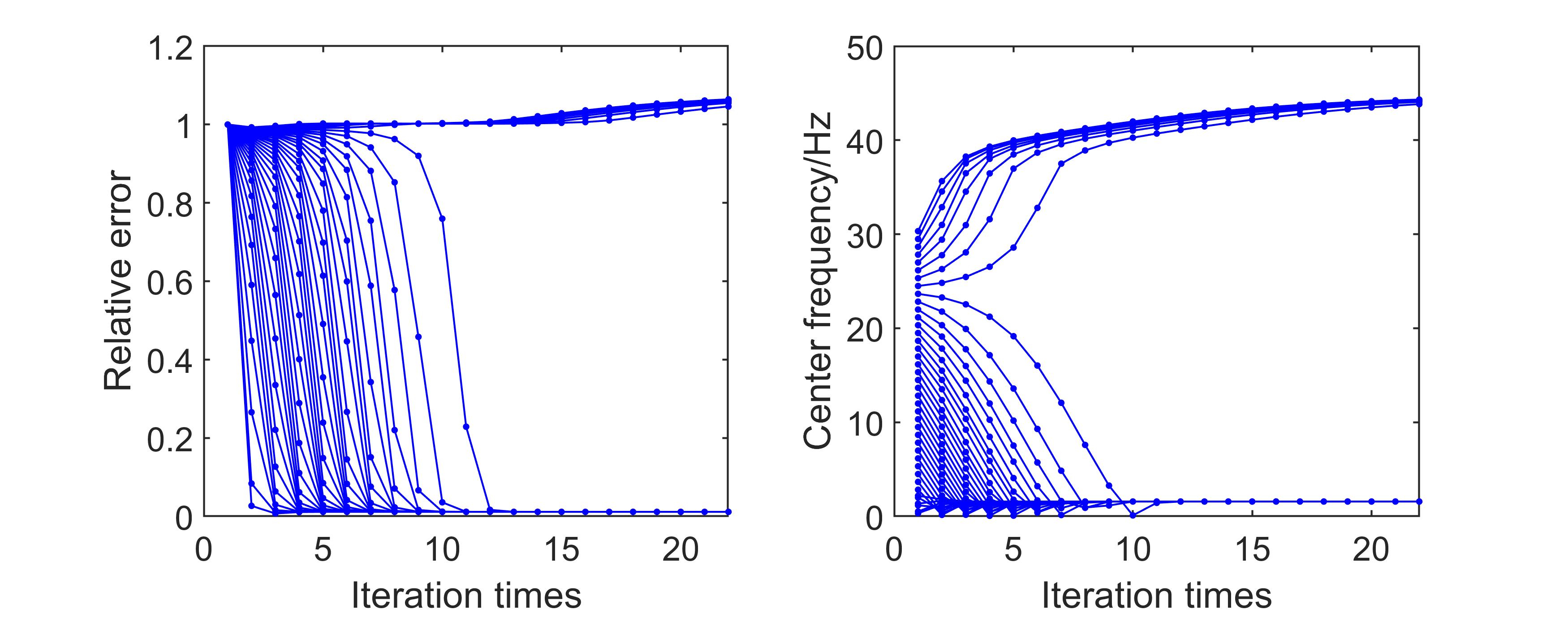}}
	\subfigure[Convergence for C2]{\includegraphics[width=3.5in]{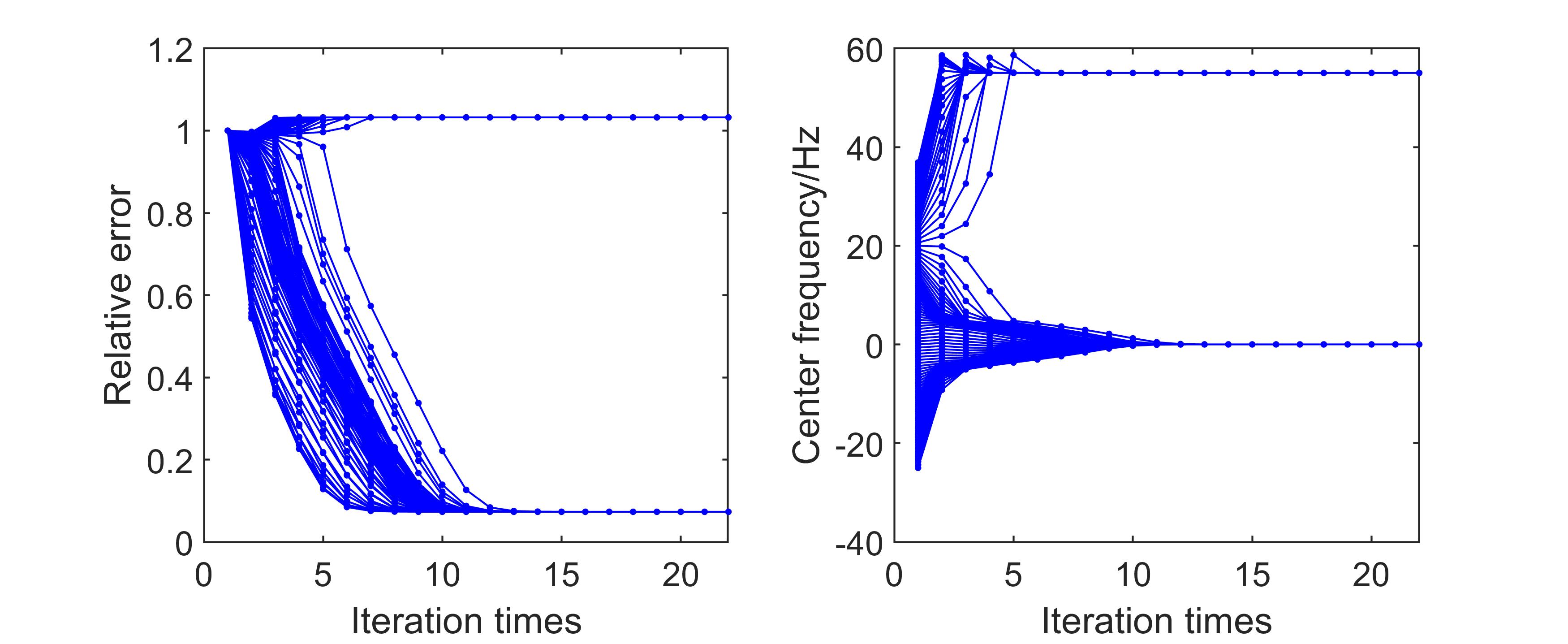}}
	\subfigure[Convergence for C3]{\includegraphics[width=3.5in]{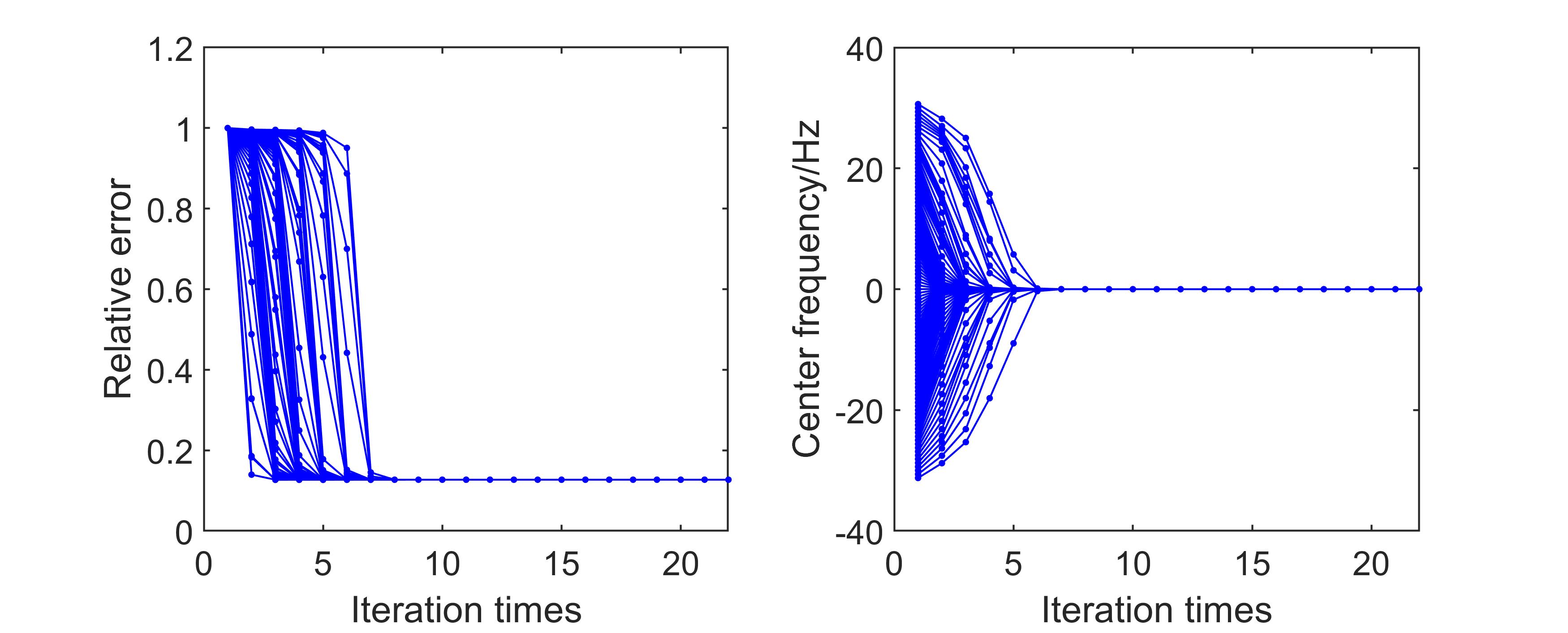}}
	\caption{Convergence test under different initializations}
	\label{INI_CONV}
\end{figure}

To test the convergence under different initializations, we scan the neighborhood of center frequency for each component and set a random initial value, then start the extraction process and record their results. Two parameters are used to evaluate the decomposing quality: one is the relative error between the original components and the recovered ones, shown in \eqref{ER}; the other is the deviation of center frequency, noted as $D_{c}=\omega_{c,i}-\omega_{c}$. We trace the two parameters as the iteration progresses, the trajectories for each component are shown in Fig. \ref{INI_CONV}.
\begin{equation}
{\rm{ER}}(i,k)=\frac{|u^{k}_{i}-f_{i}|}{|f_{i}|}
\label{ER}
\end{equation}
In this equation, ${\rm{ER}}(i,k)$ is the deviation of $i^{th}$ component at $k^{th}$ iteration, $f_{i}$ is the $i^{th}$ true mode, $u^{k}_{i}$ is the corresponding recovered component. 

In Fig. \ref{INI_CONV}, the vertical axis in the left subfigure is relative error ER, in right is $D_{c}$  with unit Hz; meanwhile, the abscissa axis is iteration times. For the trend line, noted as C1, we can see that to promise convergence to the correct mode, the initial location can deviate from the true center frequency up to 25Hz, around at middle point between center frequencies of C1 and C2. However, beyond such range, it would converge to an aliased mode. Depending on the initial position, it takes at most 12 times to reach the limit. Undoubtedly, closer to the true center frequency, less time is needed to converge. To evaluate the convergence speed and accuracy, we also plot the ER graphs, arranged in the left subfigures. We can see in the convergent region, the relative error ER would finally reduce to $0.9\%$.

The decomposition for C2 and C3 shows similar behaviors. Due to the absence of extracted components, the convergence interval is broader, and thus convergence is much easier to achieve. The final ER value is about $6.5\%$ for C2, $12\%$ for C3. Besides, what is worthy to mention, it is also observed that our algorithm is insensitive to the initial amplitude selection. So, a recommended fast initialization approach is firstly to search the peaks, then do the ranking for them in descending order, finally select one by one from the sorted peak series as the initial values for each decomposing process respectively.

\begin{figure}
	\centering
	\subfigure[ER comparison]{\includegraphics[width=2.5in]{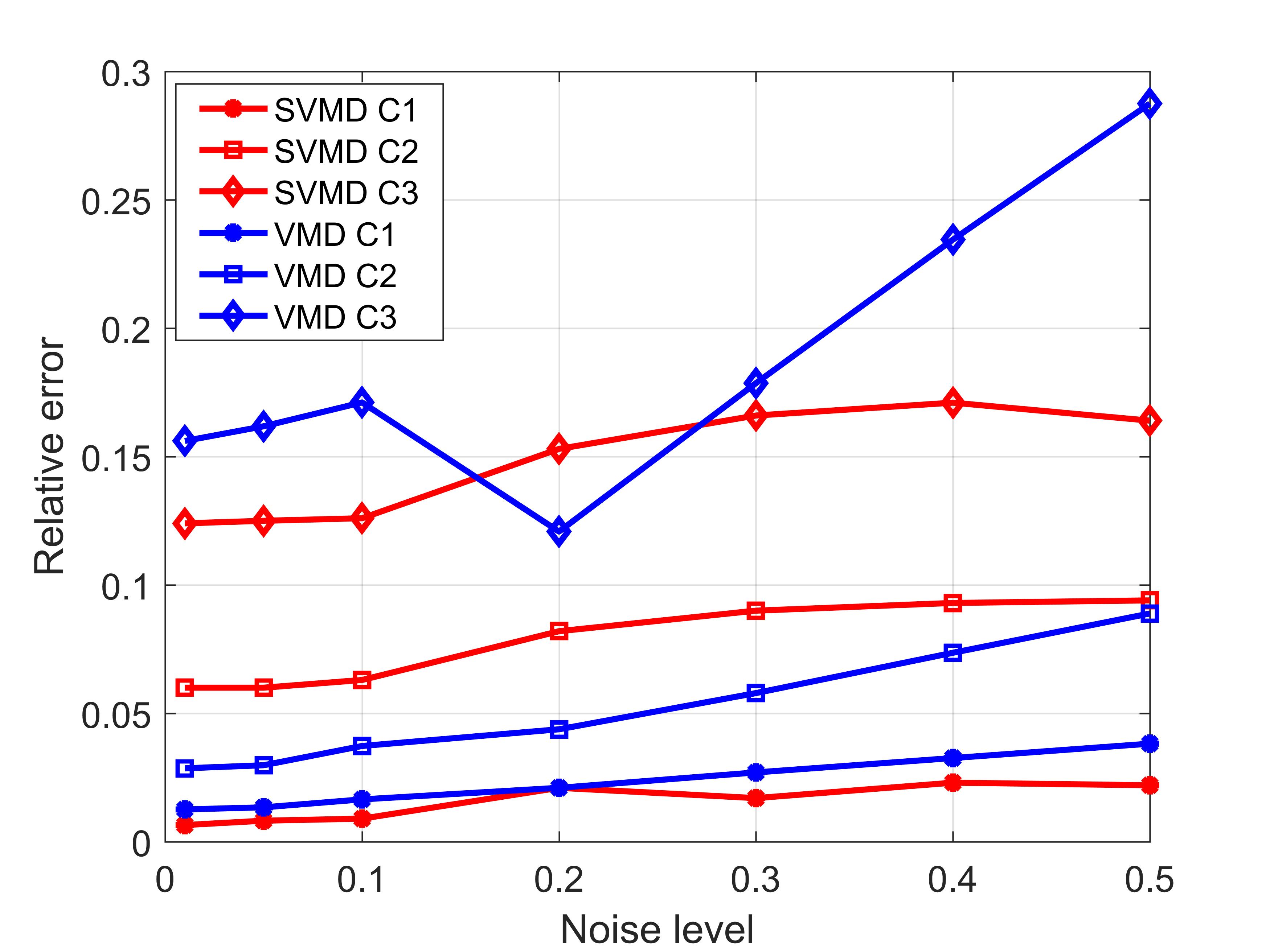}}
	\subfigure[EM comparison]{\includegraphics[width=2.5in]{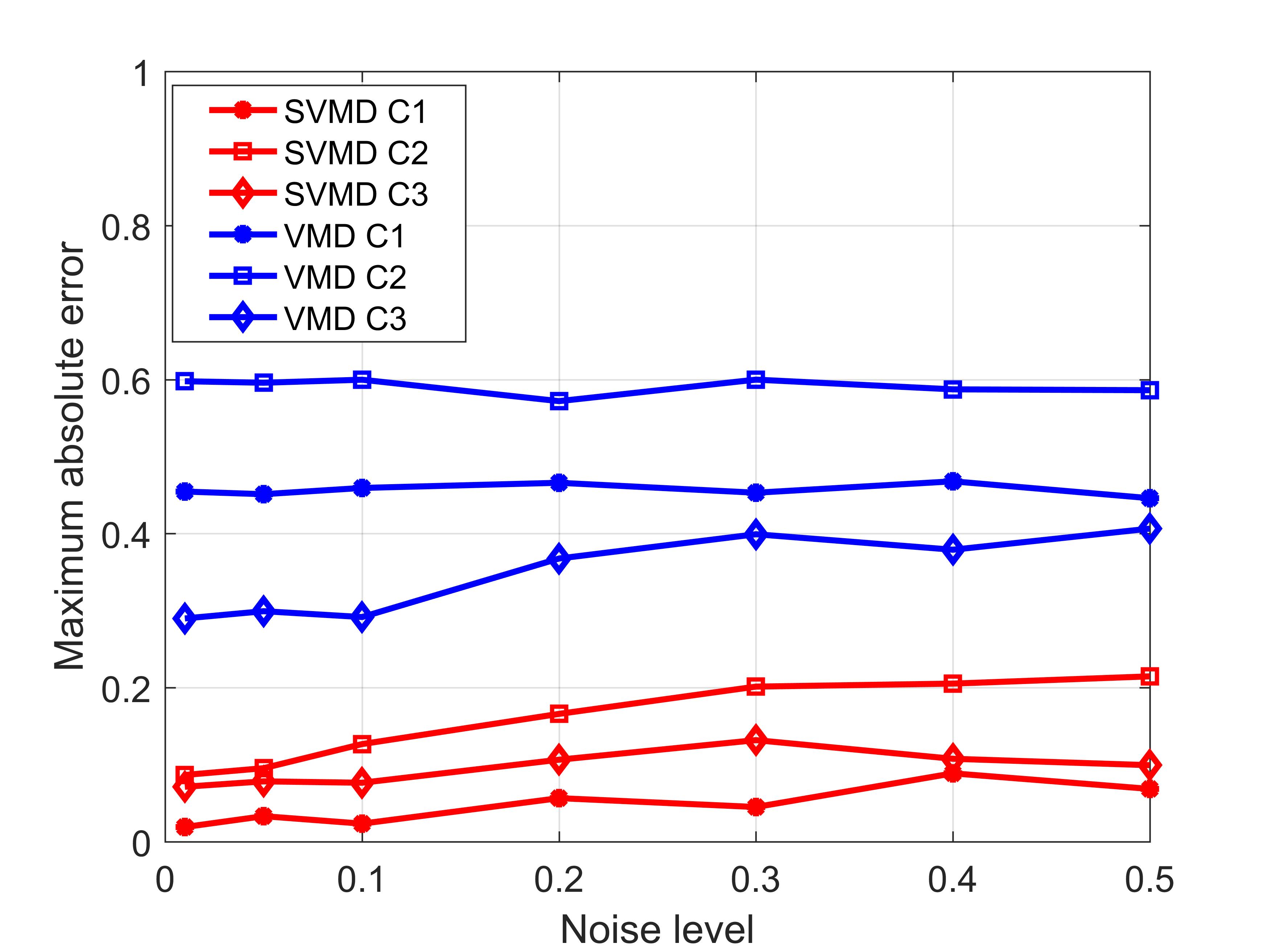}}
	\caption{Noise susceptibility}
	\label{NoiseSus}
\end{figure}

In addition, the noise level is also a critical factor that affects the algorithm performance. Convergent range and limit accuracy would change as the variation of noise power. To evaluate this influence, we tune $\sigma$ from 0.01 to 0.5, and evaluate the corresponding convergence behaviors, such as convergent interval, maximum error (EM), and average relative error (ER). EM is the maximum absolute difference between the extracted components and true modes. Relevant results are listed in Table \ref{CONVINT} and exhibited in Fig .\ref{NoiseSus}.

\begin{table}[h]
	\centering
	\caption{Convergent interval at different noise levels}\label{CONVINT}
\begin{tabular}{| c | c | c |}
	\hline
	 Noise level & C1(center=2Hz) & C2(center=45Hz)\\ 
	\hline
	 0.01 & [-2,15] & [-44,6]\\
	\hline
	 0.05 & [-2,18] & [-44,8]\\
	\hline
	 0.1 & [-2,22] & [-37,20]\\
	\hline
     0.2 & [-2,33] & [-38,43]\\
    \hline
     0.3 & [-2,28] & [-43,43]\\
    \hline
     0.4 & [-2,26] & [-43,42]\\
    \hline
     0.5 & [-2,24] & [-43,40]\\
	\hline
\end{tabular}
\end{table}

Combined Table \ref{CONVINT} and Fig. \ref{NoiseSus}, it implies that proper strong noise addition can widen the convergence interval, but without inducing extraction quality worsening greatly in our algorithm. It is an interesting phenomenon that also has been observed in \cite{VMD}. A possible reason is that higher background noise can lower the gap thus help to escape the trap of wrong local extremes. 
From the results shown in Fig .\ref{NoiseSus}, it can be observed that, excluding ER of C2, SVMD performs better than VMD, especially for EM reduction. By the way, we also conducted the test for the EMD method, but its performance is badly worse than these of SVMD and VMD, so not presented here. 

\subsection{Refinement and mode number determination}
\begin{figure}
	\centering
	\includegraphics[width=2.5in]{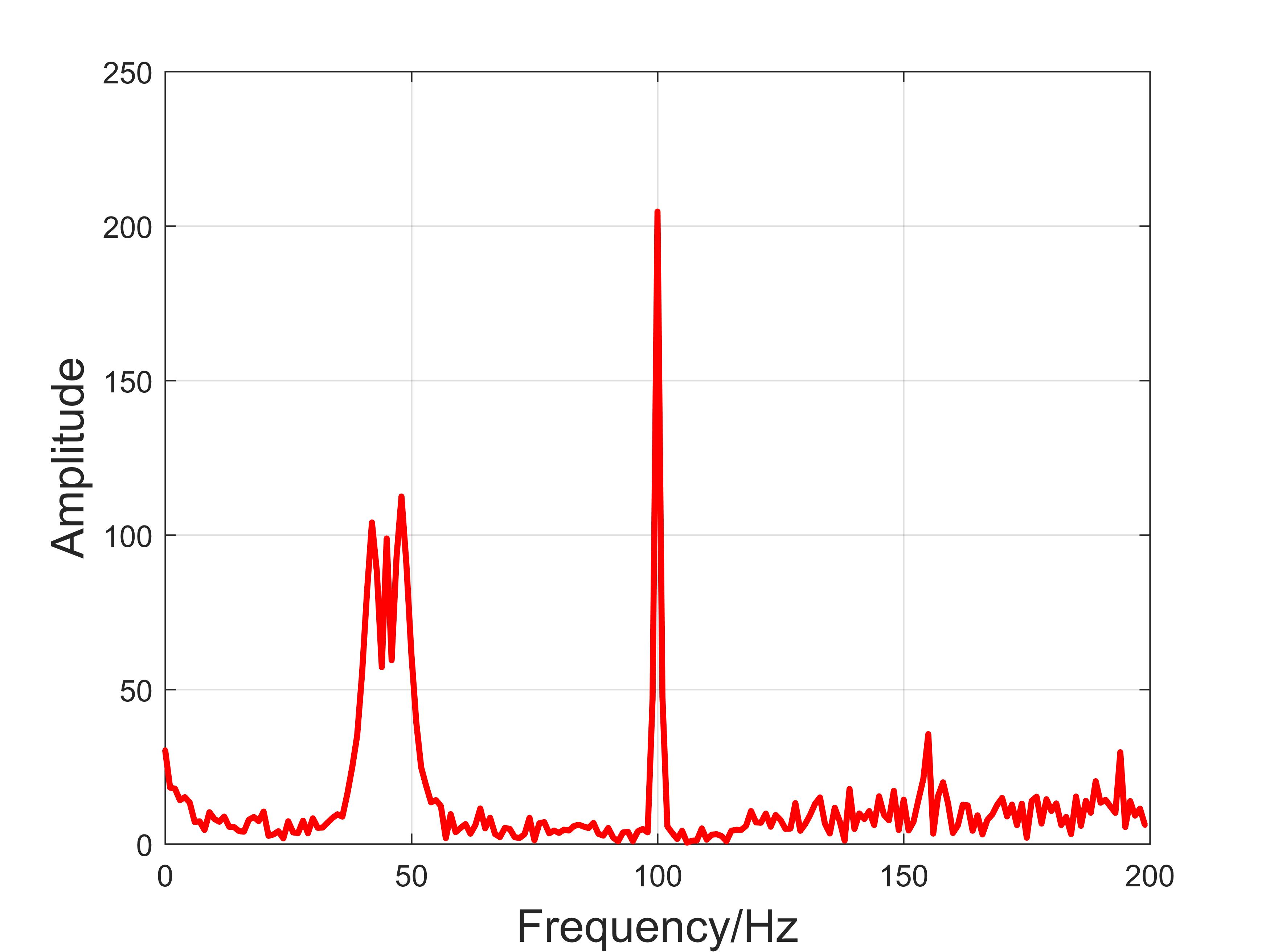}
	\caption{Spectrum after cycle I}
	\label{Spectrum1Cycle}
\end{figure}

From the spectrum obtained after the first three times of extraction shown in Fig .\ref{Spectrum1Cycle}, it is observed that there are still several components with a similar profile as previous ones but of low power remaining in the same region. These residuals can compensate for the deviations between obtained components and true ones. So to improve the accuracy, we can repeat the decomposition operation to the residual spectrum and merge components with similar center frequency together. To verify the feasibility of this operation, we conduct a test using again signal 1 with noise level $\sigma$=0.1.

Differing from the previous decomposition procedure, to extract the small residual components, we must lower the outer stop threshold $\epsilon$ to explore more details of the target spectrum. In fact, we can repeat this operation for several times, but because of the existence of noise, further improvement for the decomposing accuracy is limited. Over-extracting will be helpless for higher accuracy but oppositely bring more error. So the refinement should be carefully arranged and a proper stop criterion must be set, the power ratio between the concerned region and background noise is a reasonable option.

\begin{figure}
	\centering
	\includegraphics[width=2.5in]{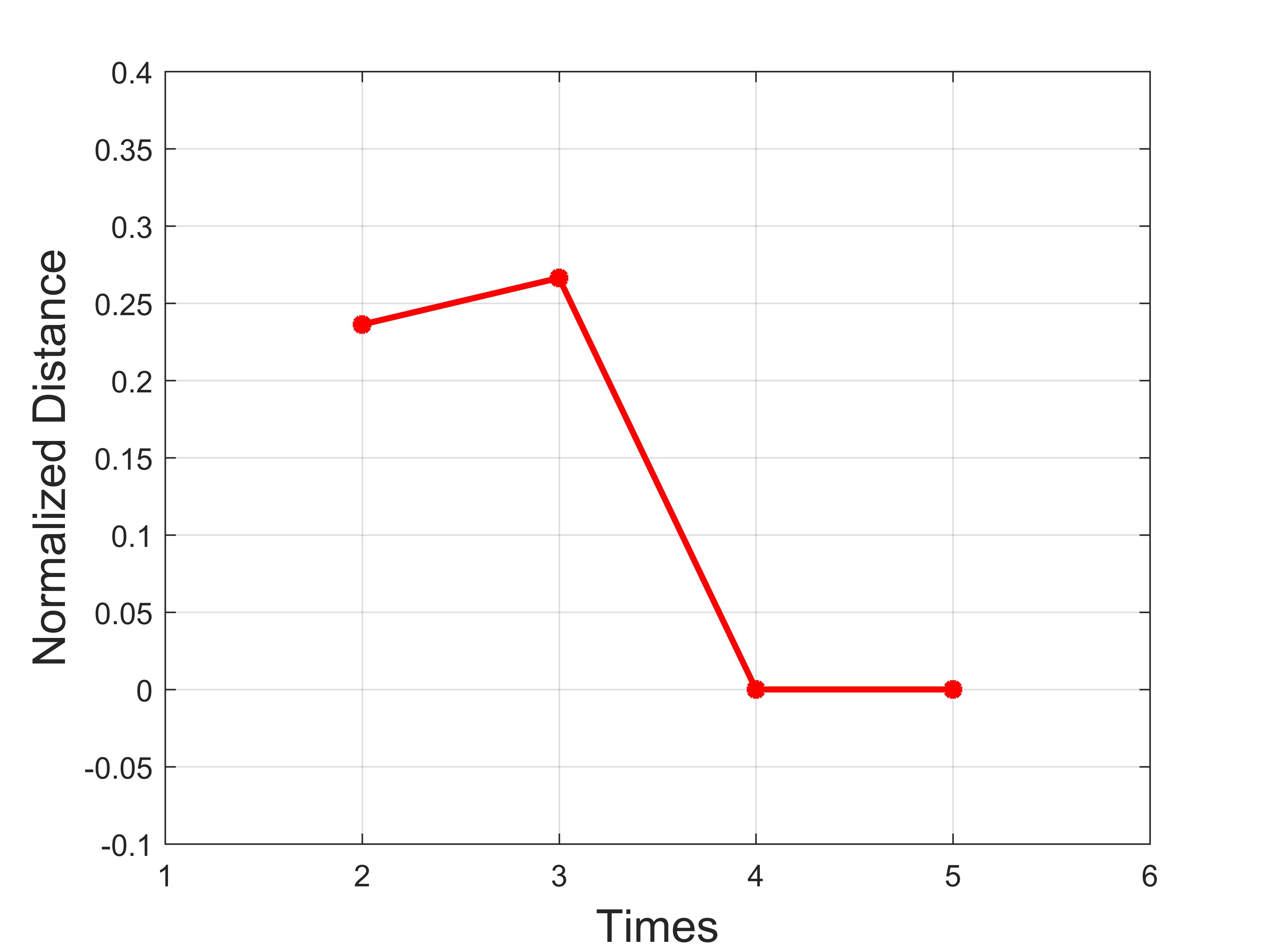}
	\caption{Mode number determination}
	\label{ModeDeter}
\end{figure}

As mentioned before, the residual components will be similar with the previous ones, so by monitoring the similarity among the obtained components, we can deduce the mode number. Here we use the minimum value of normalized distances between the current component and the previous ones as the critical parameter, when it sharply jumps to a near-zero value, the mode number is determined. Fig. \ref{ModeDeter} shows the procedure for signal 1, we can see that the mode number of this signal is 3.

\begin{figure}
	\centering
	\includegraphics[width=3.5in]{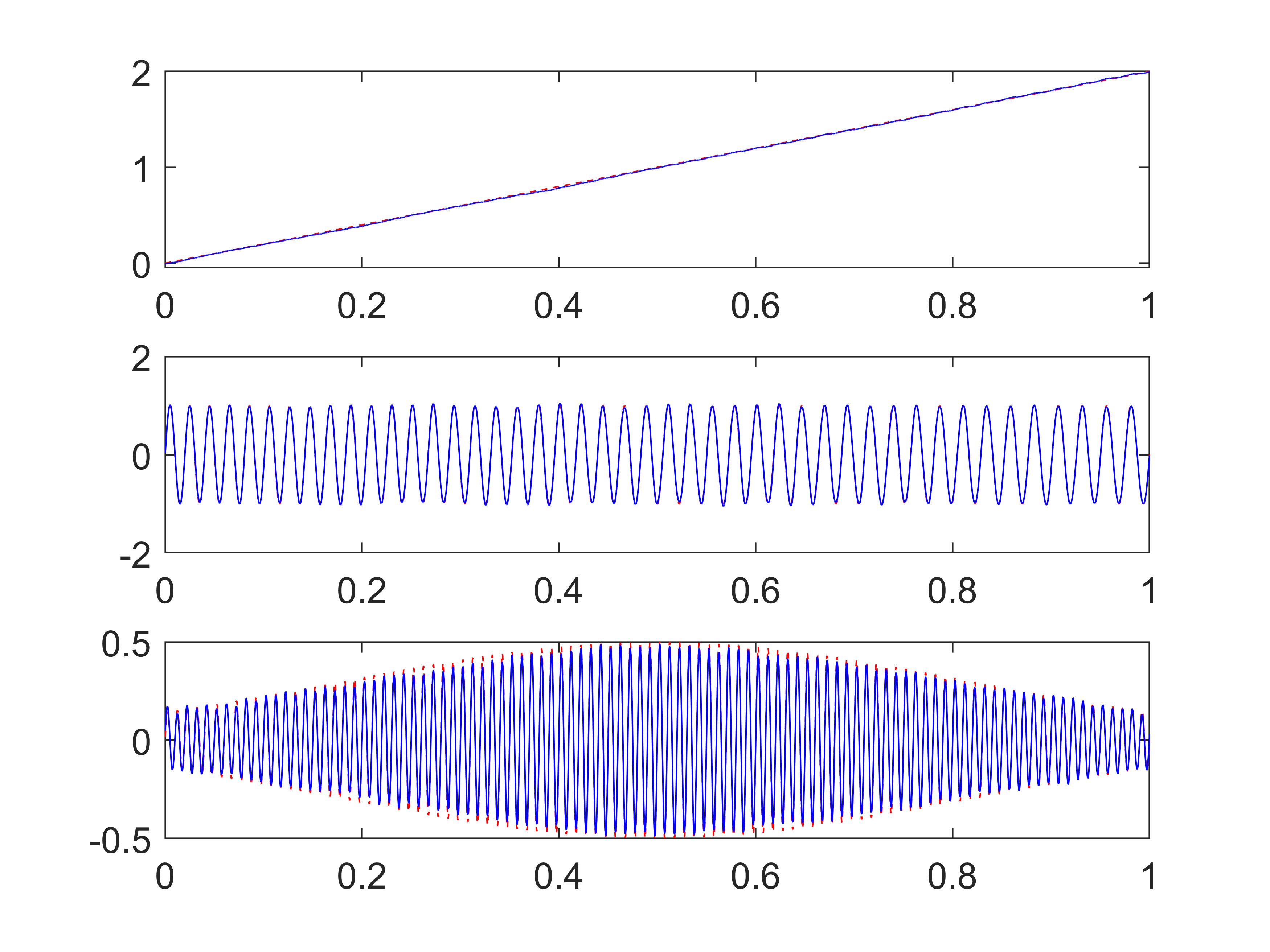}
	\caption{Extracted components after once refinement }
	\label{REFI}
\end{figure}

Since in this case, there are three components in the mixture signal, we can call three times of extraction "one cycle". Fig. \ref{Spectrum1Cycle} shows the spectrum after the first cycle, which is similar to the original one in Fig. \ref{Spectrum1}, there are three peaks remained at about 0Hz, 45Hz and 100Hz, which corresponds to the three principal components respectively. But it also is noticed that the peak of C1 is rather insignificant. To avoid divergence, we can just use these three locations and associated amplitudes as the initializations for the following optimization process. And refining results are shown in Fig. \ref{REFI} and Table \ref{REFTABLE}. 
\begin{table}[h]
	\centering
	\caption{Performance comparison of refinement}\label{REFTABLE}
	\begin{tabular}{| c | c | c | c | c |}
		\hline
		 Component&\multicolumn{2}{c|}{Before}& \multicolumn{2}{c|}{After} \\
		\cline{2-5}
		\multicolumn{1}{|c|}{}&ER&EM&ER&EM\\
		\hline
	     C1&0.0105&0.0275&0.0060&0.0242\\
		\hline
		 C2&0.0696&0.1232&0.0293&0.0725\\
		\hline
		 C3&0.1381 & 0.0897 & 0.0723 & 0.0535\\
	   \hline
	\end{tabular}
\end{table} \\

From Table \ref{REFTABLE}, we can see that once refinement for signal 1 with noise level 0.1 can nearly halve both ER and EM even from such a low level, which is a rather excellent advancement compared with VMD and EMD. This approach allows us to obtain more accurate results without much time cost, since as mentioned before, each iteration number in SVMD is very small. Fig. \ref{REFI} shows the final obtained components and the true modes, we can observe that their curves almost perfectly coincide, only in C3 there are some tiny visible deviations. 

This technique is not so easy or convenient to realize for the EMD method. Firstly, the EMD algorithm would never consider the frequency domain, refinement will be hindered primarily. Tough, it seems we can adopt a similar approach for EMD: transform the extracted signals into the frequency spectrum, then add components with the close center frequency together. However, the main issue is the output order. In EMD, modes are extracted in decreasing order of frequency. Unfortunately, these modes in the front queue are commonly unwanted high-frequency noise. To refine the results, we must wait until all the components have been extracted, which is inconvenient and low efficient. Another problem is the modal aliasing. Components extracted by EMD are always mixed with each other, post-processing should pay more attention to separate them, only emerging operation may make no sense.

\begin{figure}
	\centering
	\subfigure[VMD with K=2]{\includegraphics[width=3.5in]{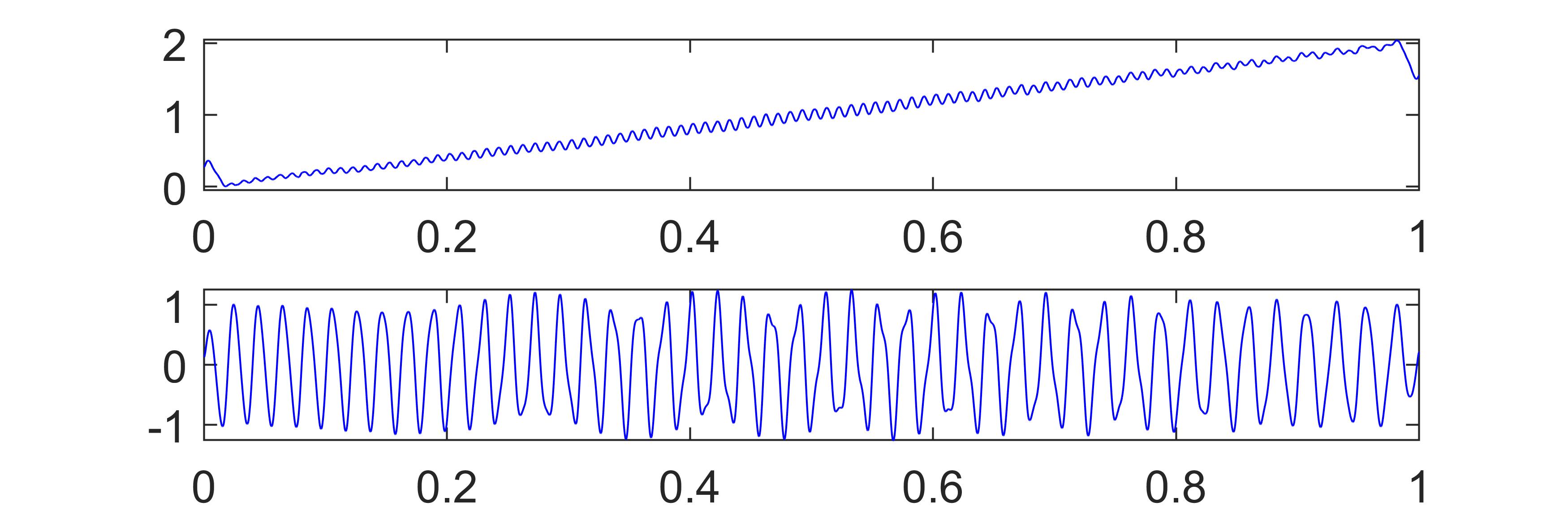}}
	\subfigure[VMD with K=5]{\includegraphics[width=3.5in]{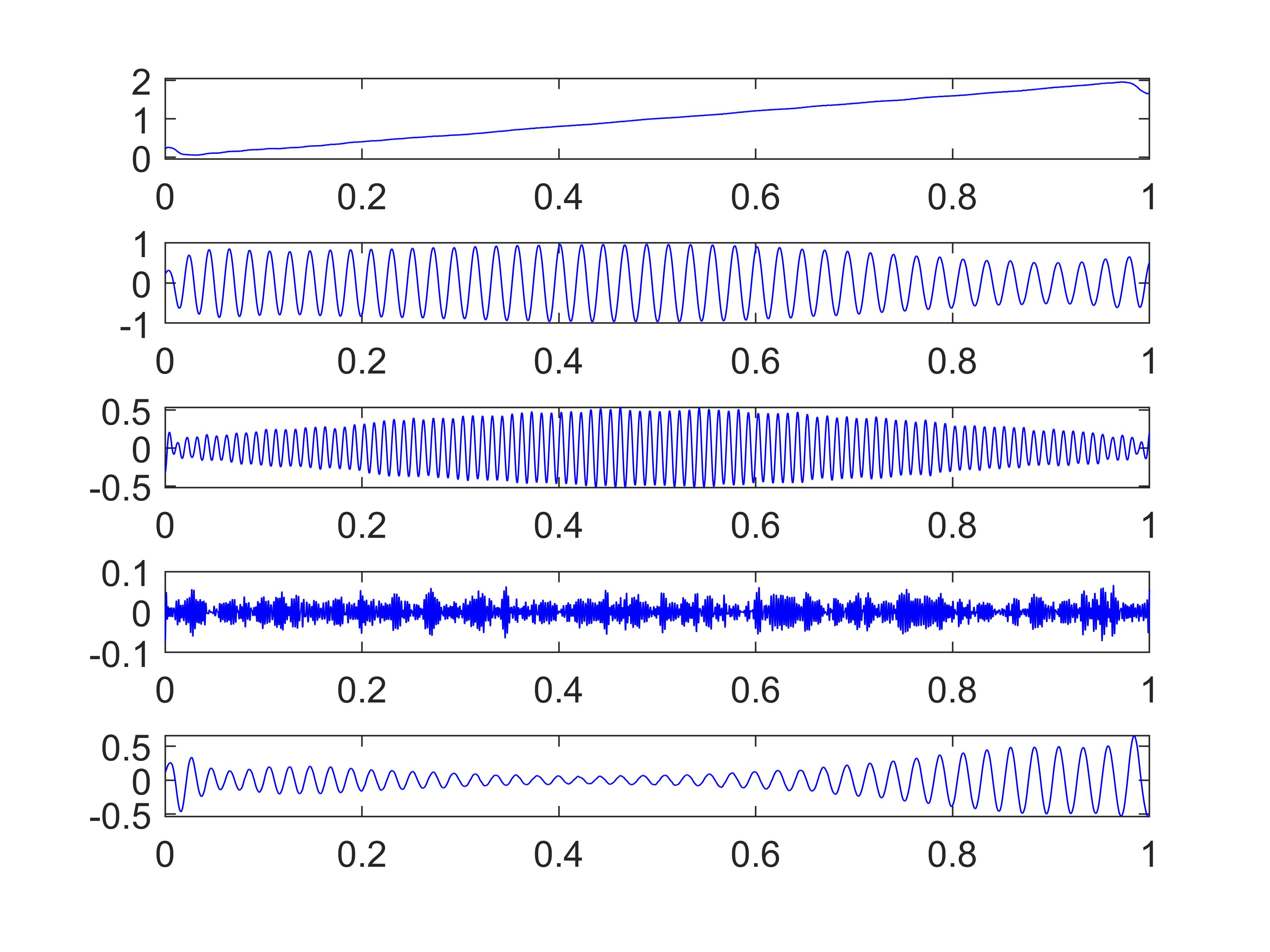}}
	\caption{VMD algorithm with different assumptions in mode numbers}
	\label{VMDwithK}
\end{figure}

Apparently, the same refinement approach should also work for VMD with a correct pre-defined number of modes. But from our tests, it works accidentally: for some cases, the results show some improvement, while sometimes even worse. There are two possible explanations: one is the profiles of the remaining components may lose fidelity with the previous ones, thus the decomposition results of the second cycle may also deviate from these of the first cycle; secondly, as shown in Fig. \ref{Spectrum1Cycle}, after the previous cycle, the current mode number may vary, thus a new assumption about it must be given. Another possible refining technique can be designed to increase the pre-defined component number and add modes with similar center frequency together. However, under assumptions with diverse numbers of modes, the VMD algorithm would generate totally different results: underbinning would certainly cause modes mixed; while overbinning, the mode with broad bandwidth will be separated into two or more modes. An example of a mixture signal described in \eqref{sig1} using VMD with different mode number assumptions is shown in Fig. \ref{VMDwithK}. Fig. \ref{VMDwithK}a is the results of underbinning, obviously, the result is wrong, aliasing happens between the modes, inducing great deviation from the true ones. Fig. \ref{VMDwithK}b is an example of overbinning. In fact, C2 and C5 belong to the same mode. However, it is difficult to make such a judgment even from the frequency domain. So this refining technique also fails.  

\subsection{Complete tests for different types of signals}
Till now we have finished all preparation works for the signal mixture decomposition. The complete procedures can be briefly described as following: 1. Use PCR method to elongate the target signal mixture; 2. Do Hilbert transform and Fourier transform to the elongated signal, transfer signal from time domain to frequency domain; 3. Apply SVMD algorithm to the obtained frequency spectrum, extract one component, convert it back to time domain; 4. Update the signal mixture, repeat steps 1 to 3, monitoring the normalized distance to determine the number of modes; 5. Start refinement approach, sum the components belonging to the same mode together. 

To test the decomposing quality using the SVMD method, during this section, we conduct several experiments about different types of non-stationary and multimode signals with narrow bandwidth assumption to verify the feasibility of the proposed algorithm. Comparison with VMD method also would be shown, the extracted signals in the time domain obtained by using both two methods will be visualized together. Also, some parameters are proposed to evaluate the final results, including ER, EM, and end effect factor. End effect factor (noted as Q$_{ee}$), is designed to estimate the severity of end effect, since distortions in two ends would be harmful in some applications, such as time series prediction \cite{EMD_endeffect1}. It is a ratio between the average deviation of the end region and the whole region. 

In our tests, signals are all mixed with noise, which is common in real practice and also a challenge for the robustness of the algorithm. Pure harmonic signals are also rare in the following examples, since the capability of analyzing non-stationary signals covers that of stationary signals in principle. In our tests, signals with different types of trend lines and various numbers of modes would be decomposed by using both SVMD and VMD to check the generality and robustness. To explore the ultimate performance of our algorithm, the results shown are refined. 

\begin{figure}
	\centering
	\subfigure[Time domain]{\includegraphics[width=2.5in]{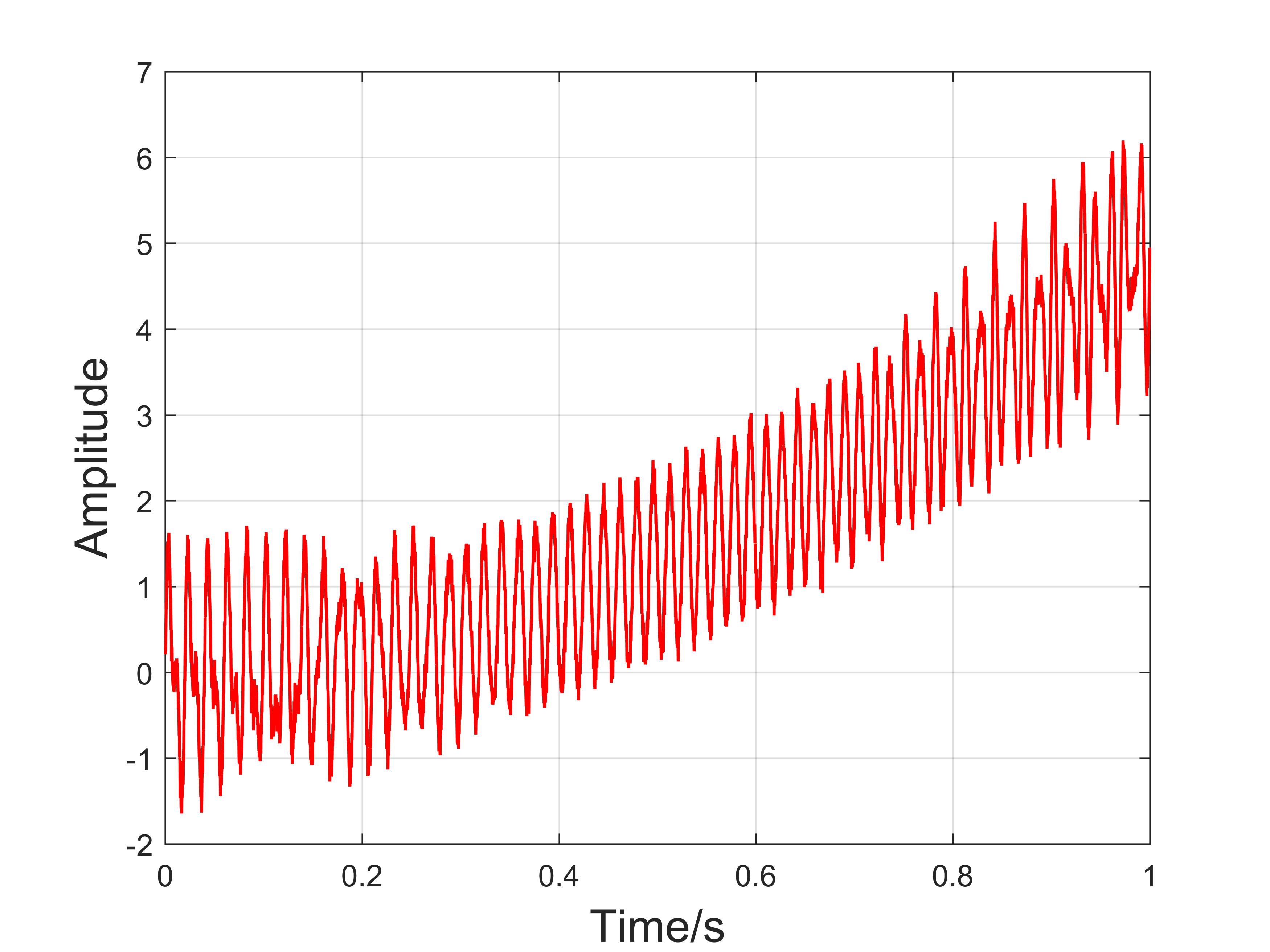}}
	\subfigure[Frequency domain]{\includegraphics[width=2.5in]{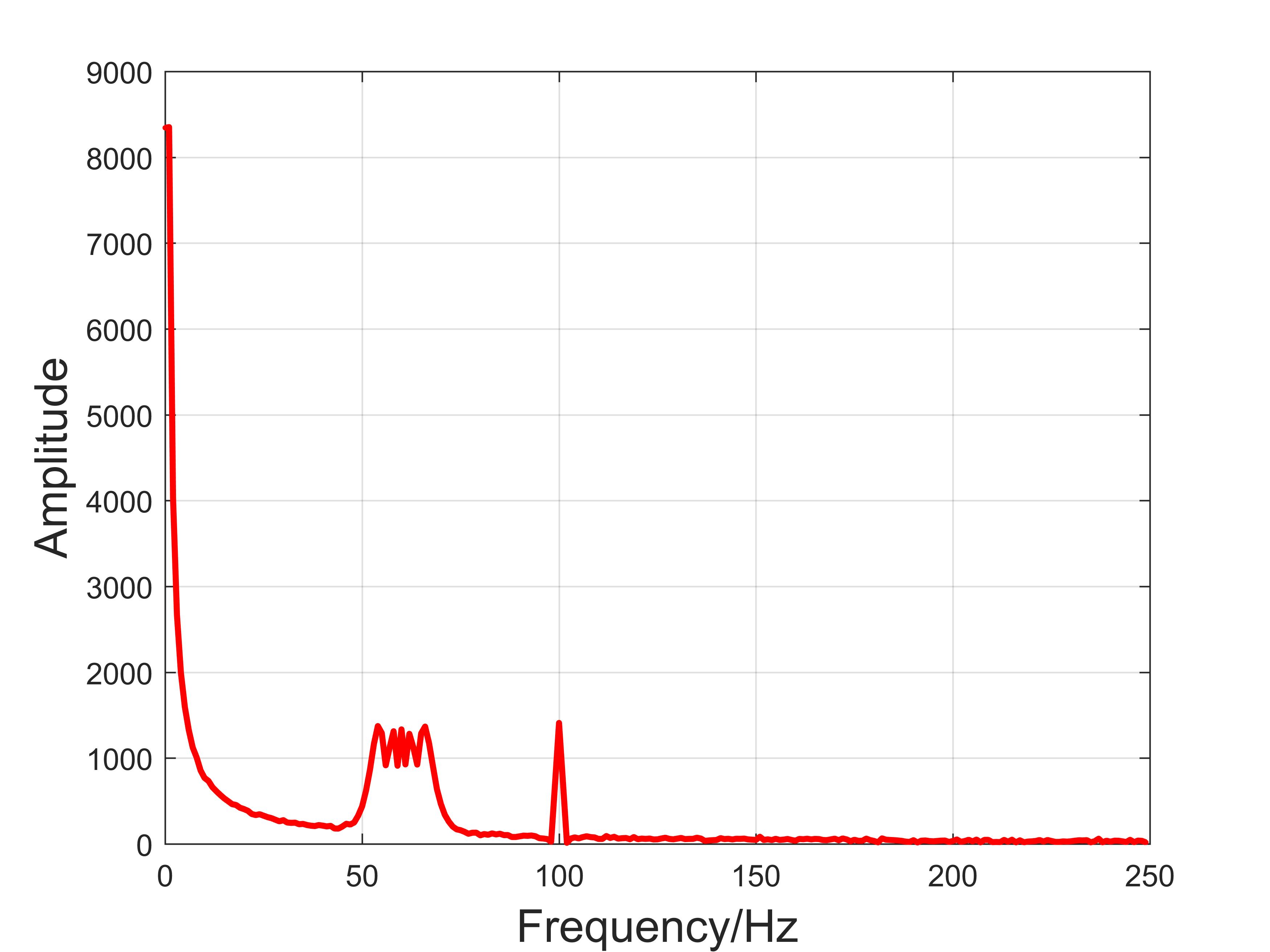}}
	\caption{Signal 2 mixed with three modes and noise }
	\label{Signal2}
\end{figure}

$(1) Example$ $ 1$: Since we has tested a signal with the trend of order one, this time we use a trend line of order 2 mixed with one FM and one AM signals:
\begin{equation}
f(t)=5t^{2}+{\rm sin}(100\pi t+20\pi t^{2})+(1-e^{-5(t-0.5)^{2}}){\rm sin}(200\pi t)+\epsilon;
\end{equation}
In our case, the standard deviation of white noise $\epsilon$ is still 0.1. The center frequencies of three modes are about 0Hz, 60Hz, and 100Hz, as Fig. \ref{Signal2}b shown. All three modes are narrowband, so it is possible to use our method to extract them one by one. The trend has high power and its spectrum spreads to a wide range, which generates moderate overlapping with the other two modes.

\begin{figure}
	\centering
	\subfigure[SVMD]{\includegraphics[width=3.5in]{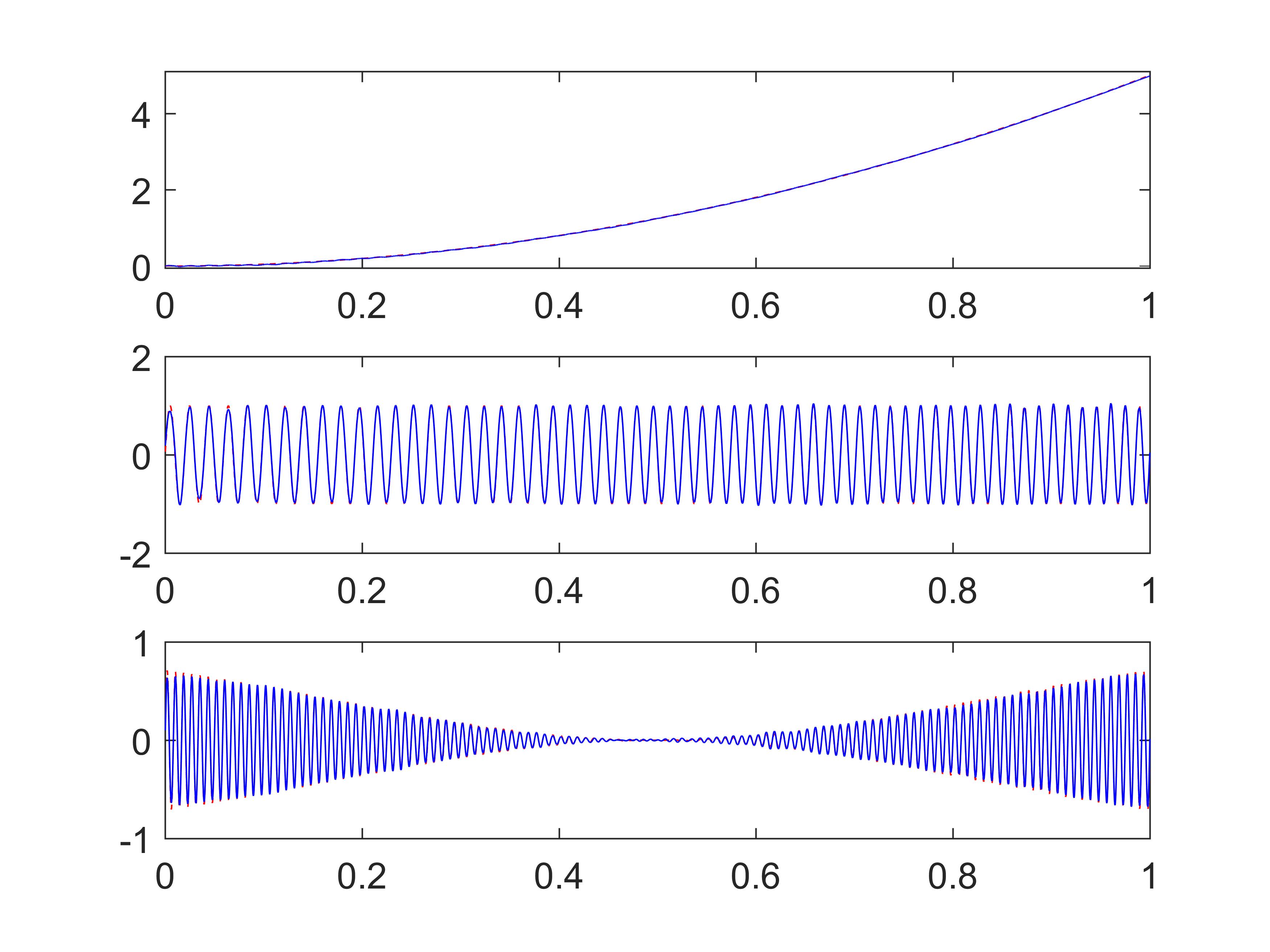}}
	\subfigure[VMD]{\includegraphics[width=3.5in]{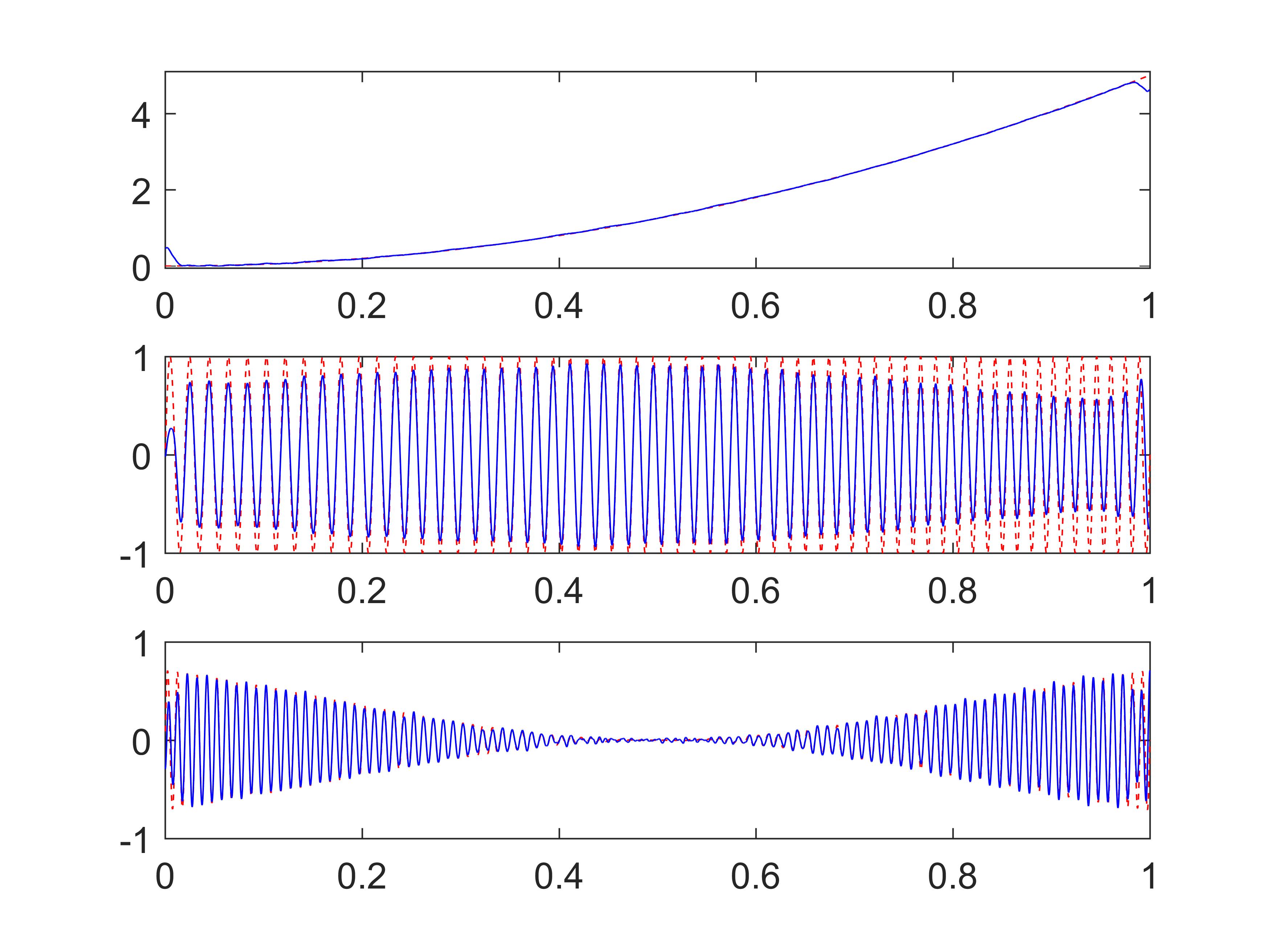}}
	\caption{Signal 2 decomposition}
	\label{Decomp_sig2}
\end{figure}
In Fig. \ref{Decomp_sig2}, the red dash line is true signal, the blue solid is recovered one. From the figure we can see that for the decomposing of signal 2, among the two algorithms, SVMD performs totally better than VMD. The extracted components precisely reconstruct the original ones, with ER $0.4\%$ and EM 0.026 for C1, ER 3$\%$ and EM 0.13 for C2, ER $5.6\%$ and EM 0.075 for C3. Only at the two ends there are slight deviations. Also the results obtained from VMD are acceptable excluding recovering C2. However, end effect still can not be neglected in C1 and C3, whose Q$_{\rm{ee}}$ parameters are 16 and 9 respectively; while for SVMD, such values are only 1.6 and 1.1. Fortunately, its influence is constrained in a short region, without extending to the middle. In VMD, other two quantitative evaluation parameters ER and EM for C1, C2 and C3 are 0.9$\%$, 22.1$\%$, 12.8$\%$ and 0.488, 0.908, 0.863, respectively. 

$(2) Example$ $ 2$: Signal 1 and 2 both ride on a monotone trend line, this time we select a valley-shaped trend. And the amplitude modulation is also more complex, with multiple periods. Finally, signal 3 is expressed as:
\begin{equation}
\begin{split}
f(t)=&5(t-0.5)^{2}+0.5{\rm sin}(50\pi t+10\pi t^{2})\\
&+0.3(1+\rm sin(5\pi t)){\rm sin}(120\pi t)+\epsilon;
\end{split}
\end{equation}

\begin{figure}
	\centering
	\subfigure[Time domain]{\includegraphics[width=2.5in]{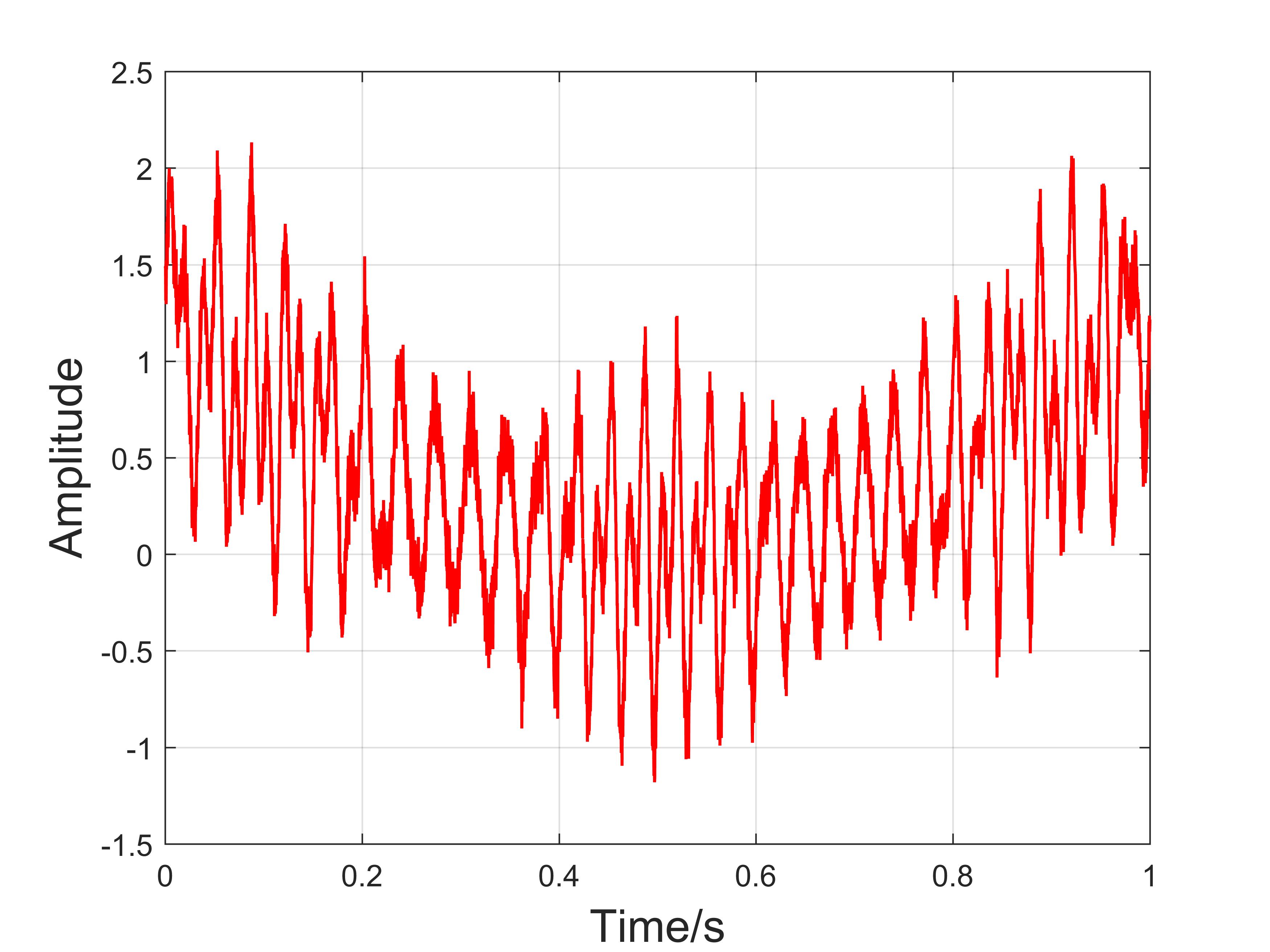}}
	\subfigure[Frequency domain]{\includegraphics[width=2.5in]{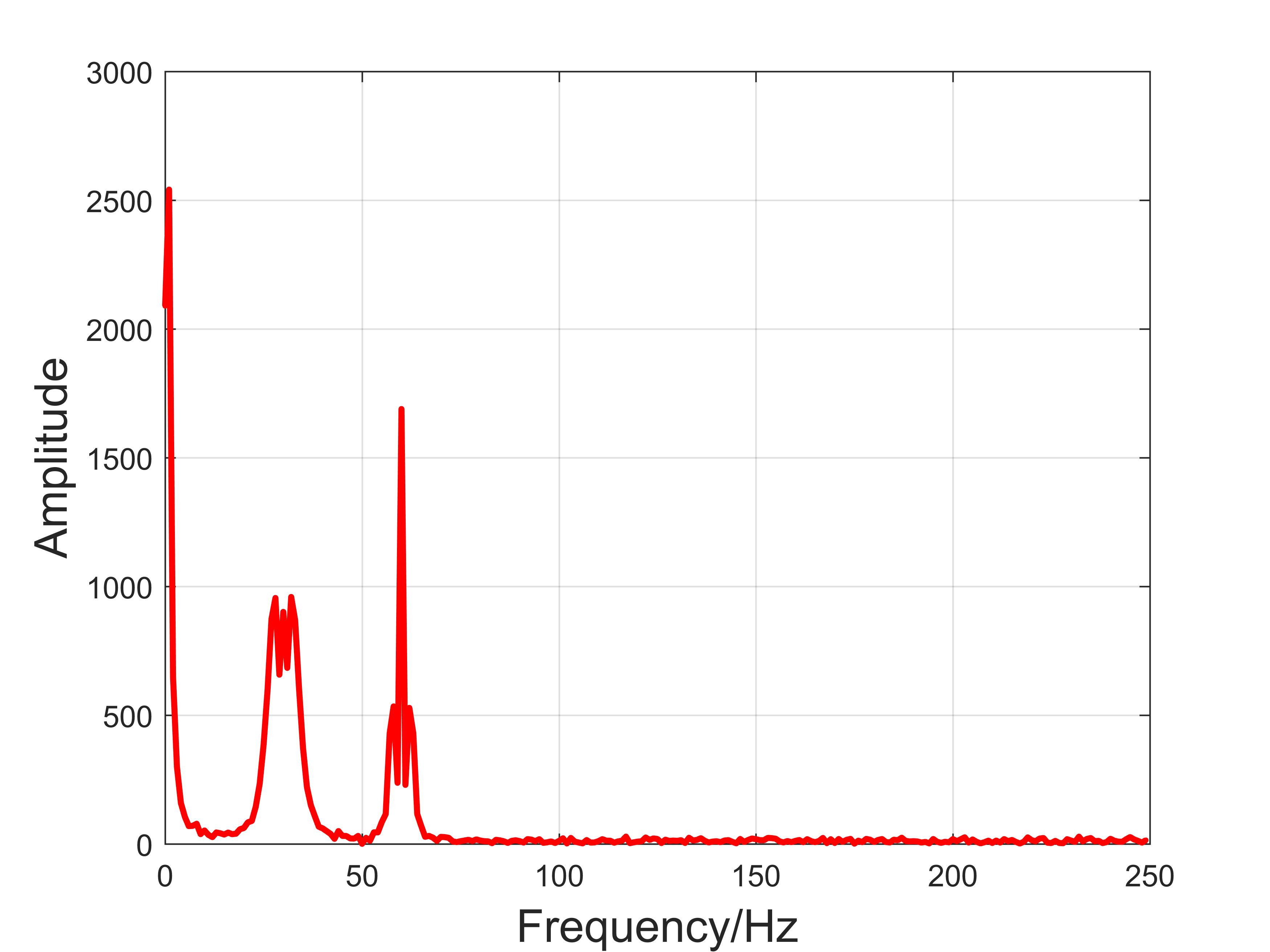}}
	\caption{Signal 3 mixed with three modes and noise }
	\label{Signal3}
\end{figure}

\begin{figure}
	\centering
	\subfigure[SVMD]{\includegraphics[width=3.5in]{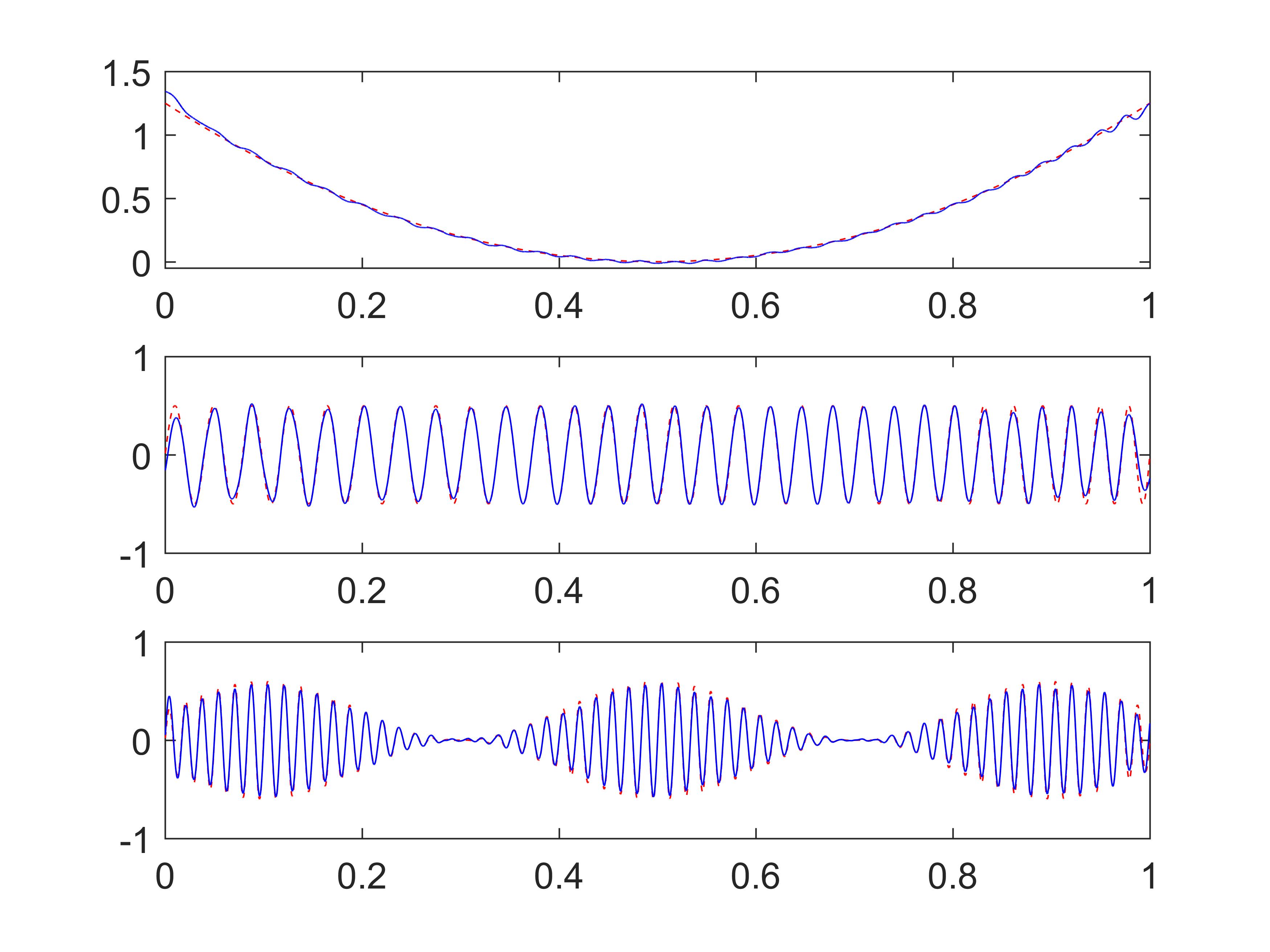}}
	\subfigure[VMD]{\includegraphics[width=3.5in]{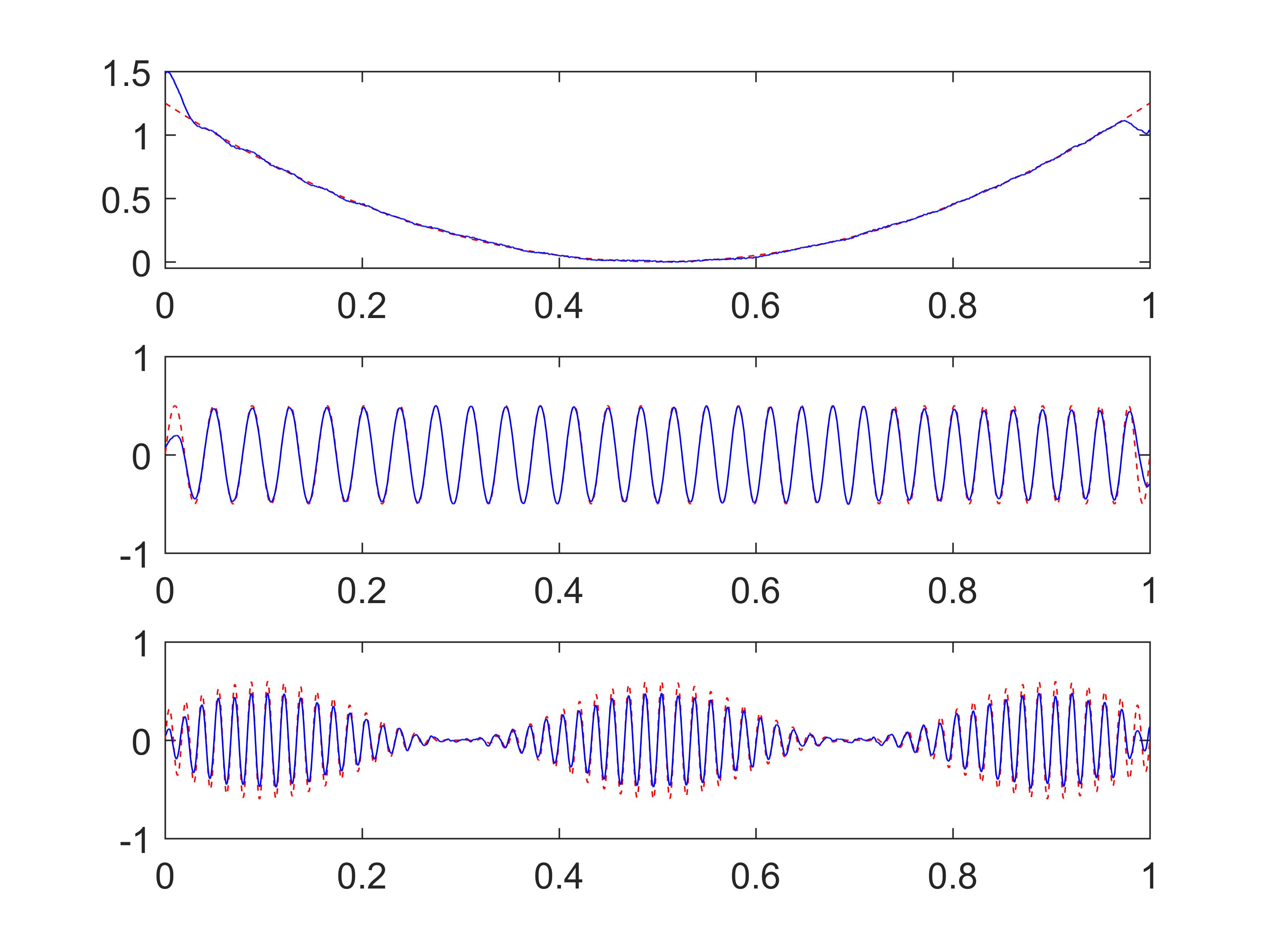}}
	\caption{Signal 3 decomposition}
	\label{Decomp_sig3}
\end{figure}

From results in Fig. \ref{Decomp_sig3}, it is easy to observe that SVMD and VMD have comparable performance. In fact, ER and EM of all three components are 2.4$\%$, 7.9$\%$, 10.1$\%$ and 0.095, 0.243, 0.177 for SVMD, 3.1$\%$, 7.5$\%$, 26.1$\%$ and 0.261, 0.358, 0.253 for VMD. Extraction for C3 using VMD is much worse than SVMD. End effect also is much stronger in VMD, for example, Q$_{\rm{ee}}$ of C1 in SVMD and VMD are 7.3 and 12.4 respectively. This example shows that the proposed algorithm has excellent capability to decompose signals with non-monotone trend and periodic amplitude modulation. These properties enhance the robustness in analyzing more complex signals and would broaden its application field such as mechanical vibration.

\begin{figure}
	\centering
	\subfigure[Time domain]{\includegraphics[width=2.5in]{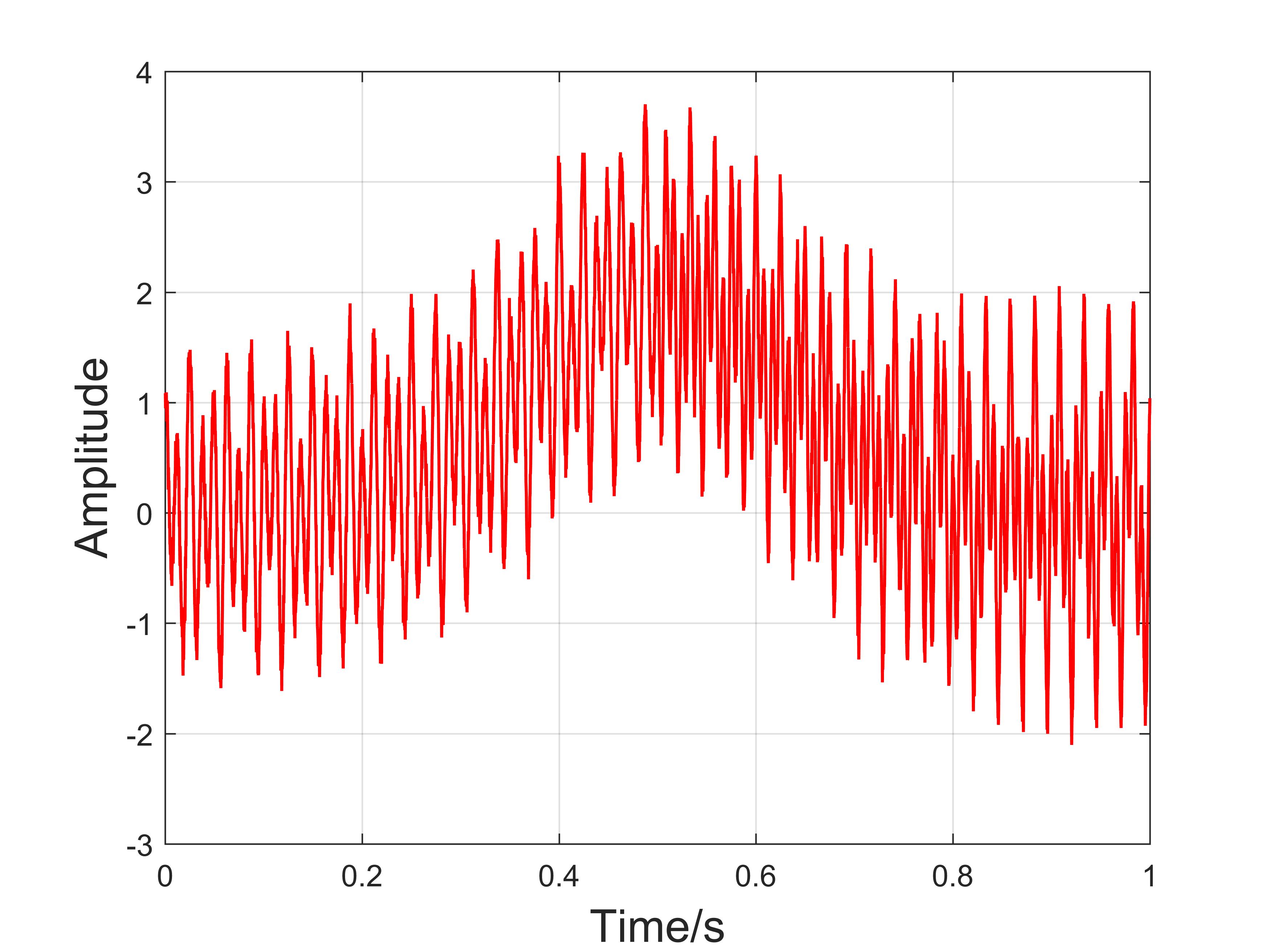}}
	\subfigure[Frequency domain]{\includegraphics[width=2.5in]{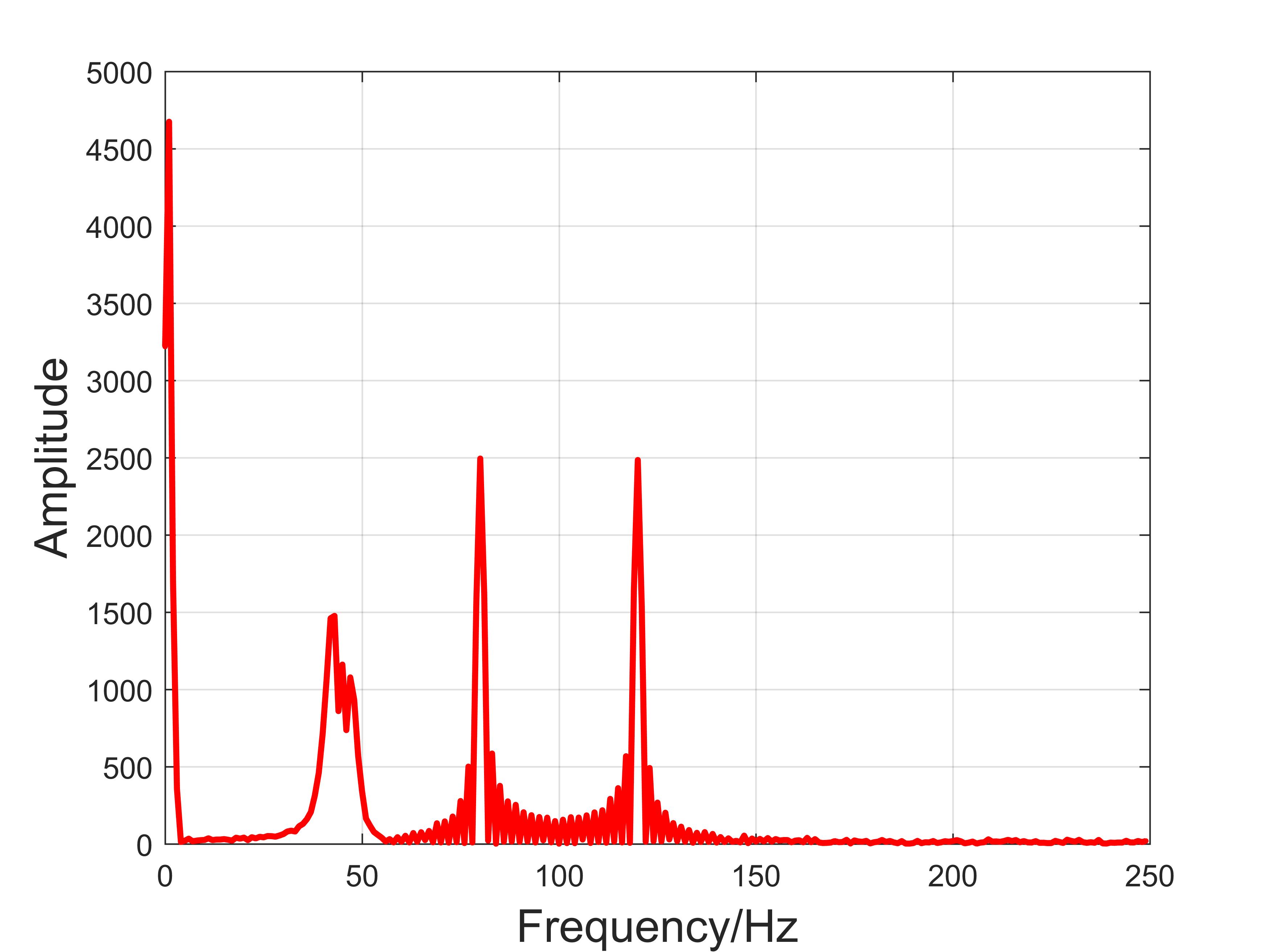}}
	\caption{Signal 4 mixed with four modes and noise}
	\label{Signal4}
\end{figure}

$(3) Example$ $3$: When the signal has no typical and simple trend such as polynomial of order 1 or 2 in the full interval, we can do the fitting for a short series of points close to the ends to obtain the trend line instead. So in this case, we firstly use a Gaussian function as the trend. Then, the second and third components are two pure harmonic signals, with each one of them only occupying half interval in time domain, but maintaining continuity at the joint. The fourth mode is a signal with both amplitude modulation and frequency modulation. The expression of signal 4 is written as:
\begin{equation}
\begin{split}
f(x)=&2e^{-30(t-0.5)^{2}}+\left\{
\begin{aligned}
\cos(160\pi t)&  &\text{0 $\leq$ t < 0.5}\\
\cos(240\pi t)&  &\text{0.5 $\leq$ t $\leq$ 1}
\end{aligned}
\right.
\\&+(0.5+0.5t^{2})\sin(100\pi t-10\pi t^{2})+\epsilon
\end{split}
\end{equation}
These complex characters of this signal obviously bring more difficulty for the mode decomposition. From its spectrum shown in Fig. \ref{Signal4}b we can see that C2 and C3 generate many high order harmonics and they spread widely from their center frequencies, thus inducing strong overlap during the region nearby. Riding on a Gaussian function also requires a fitting operation only targeted to end regions. In fact, to obtain a more precise trend line, we firstly separate the oscillating components using Fourier transform, then apply a weighted method to fit the remaining points. Considering the effect on the trend line from the samples is inversely proportional to the distance from the end, we can multiply a weighting factor which increases as points approaching the end when using "Least Square Method (LSM)" to do the fitting.

\begin{figure}
	\centering
	\subfigure[SVMD]{\includegraphics[width=3.5in]{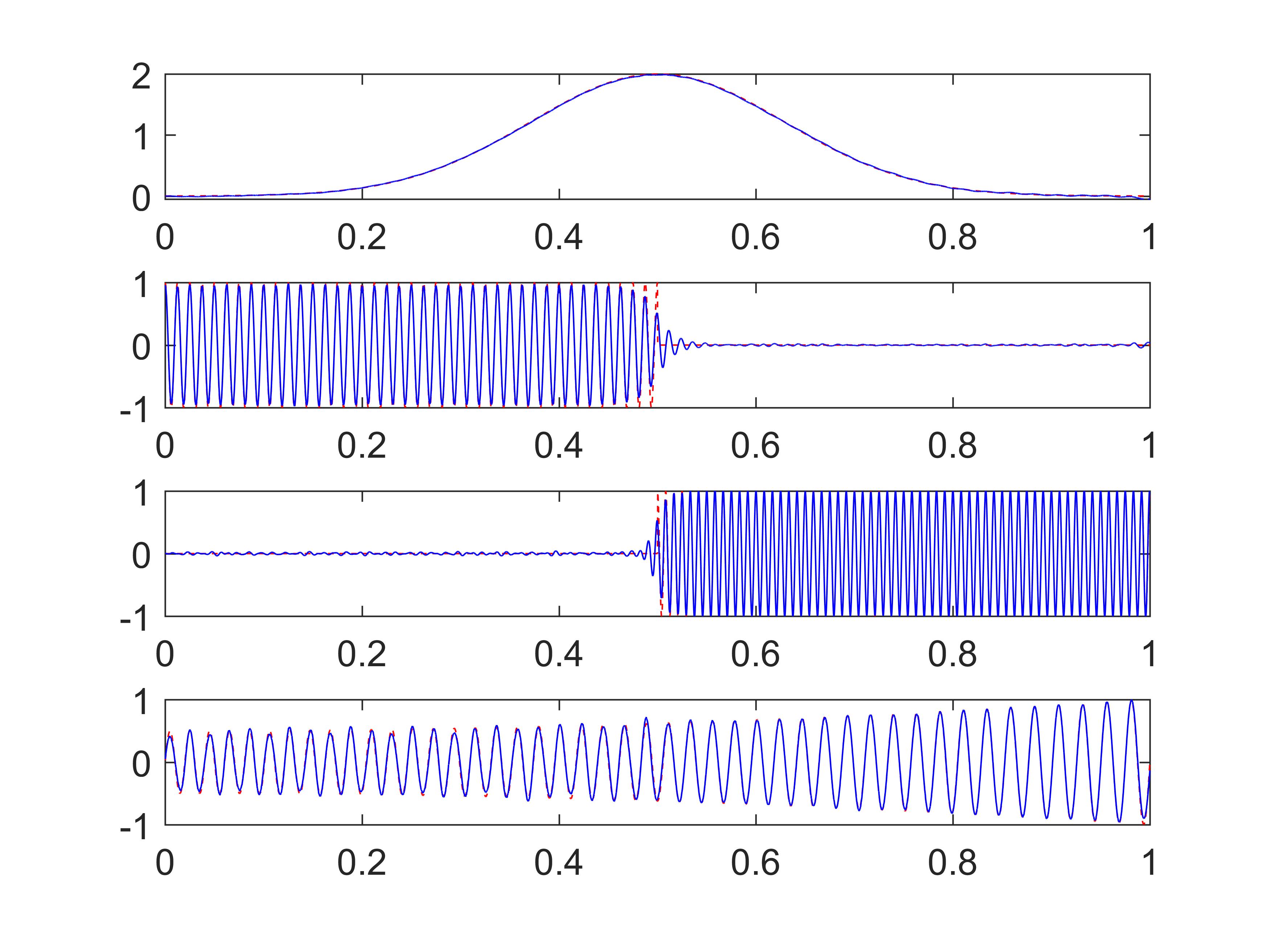}}
	\subfigure[VMD]{\includegraphics[width=3.5in]{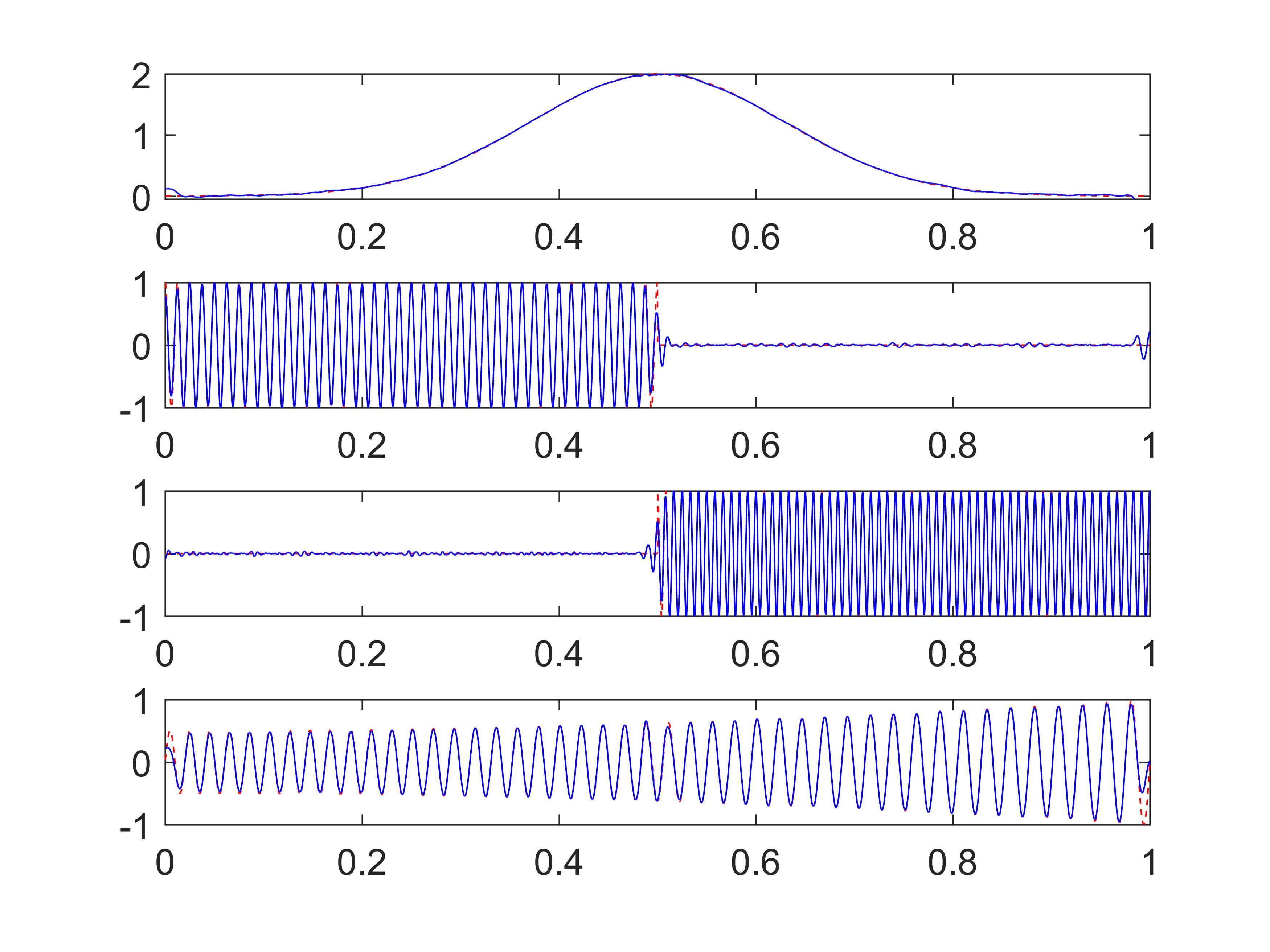}}
	\caption{Signal 4 decomposition}
	\label{Decomp_sig4}
\end{figure}

The decomposition results using the two methods are shown in Fig. \ref{Decomp_sig4}. We can see both SVMD and VMD perform excellently to extract these four modes. For C1, the corresponding ER and EM values for two methods are about 1$\%$, 0.07 and 2$\%$, 0.4, $\rm{Q_{ee}}$ shows that a stronger end effect existing in VMD. Condition is similar for C4 extraction, SVMD and VMD character with fine accuracy, ER and EM values are 6$\%$, 0.14 for SVMD, and 5$\%$, 0.6 for VMD. For C2 and C3, the results are encouraging, since their overlapping in frequency spectrum does not become a big obstacle for the decomposition, it implies a high-level capability to deal with the spectrum aliasing issue of our proposed method. 

This test verifies that our novel method is capable of processing signals containing the various number of modes and can extend the applicable field from simple polynomial trends to arbitrary ones. Also, the ability to deal with composited harmonic signals like C2 and C3 is interesting, it brings us high expectation in similar applications such as texture segmentation, crystal microstructure identification when this algorithm is extended to 2D \cite{VMD_2D}. 

\begin{figure}
	\centering
	\subfigure[Time domain]{\includegraphics[width=2.5in]{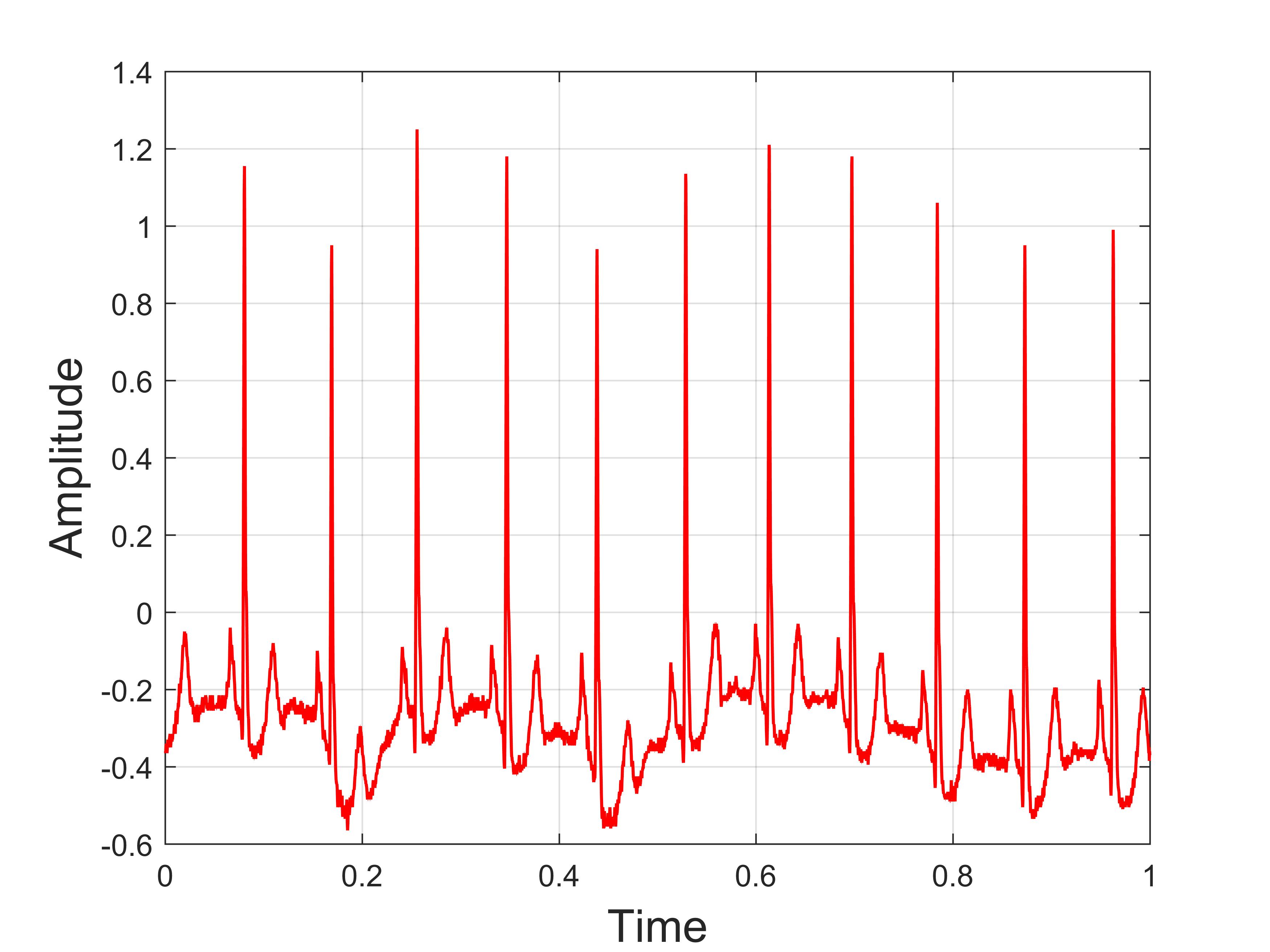}}
	\subfigure[Frequency domain]{\includegraphics[width=2.5in]{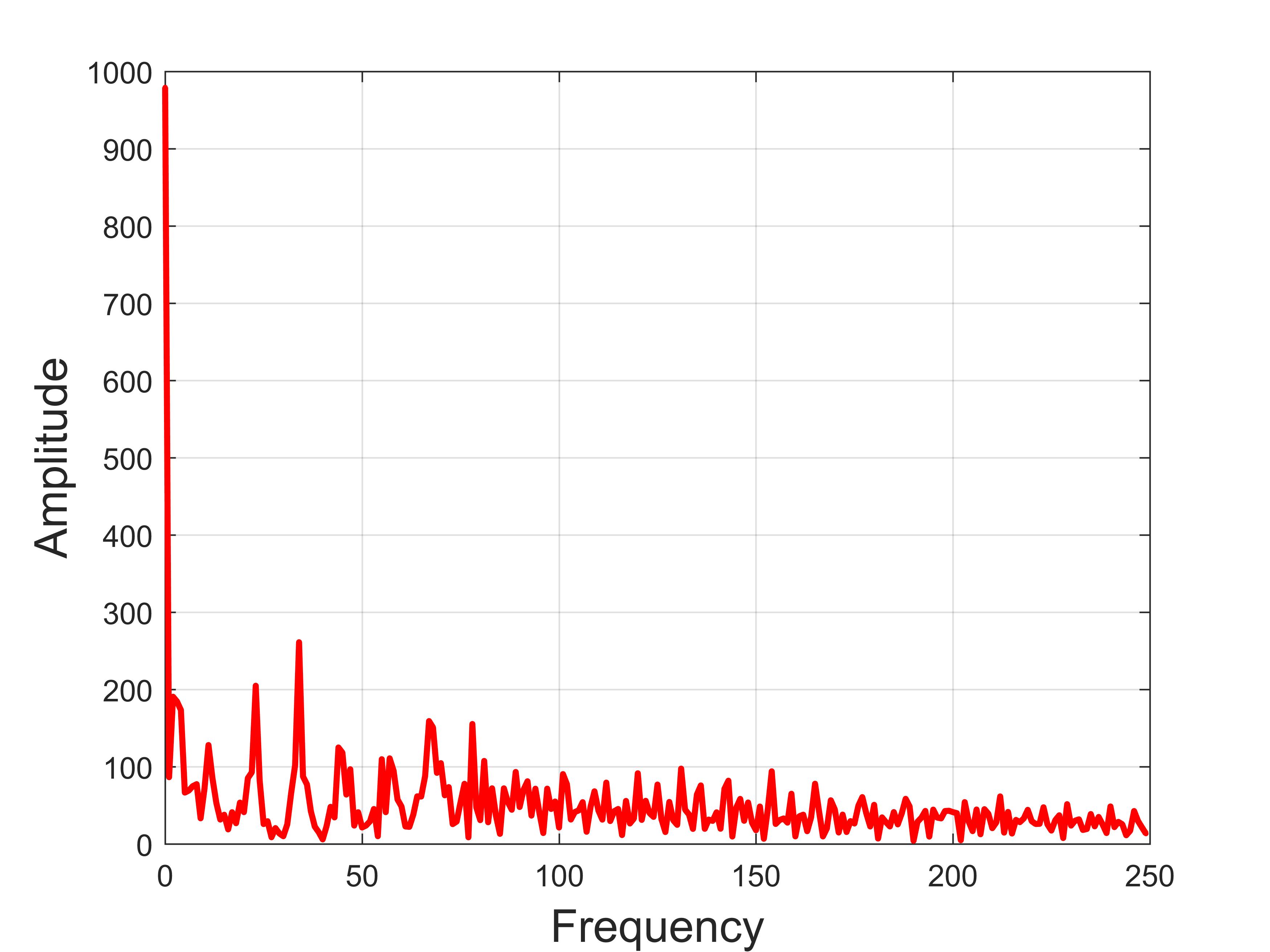}}
	\caption{ECG signal }
	\label{ECG}
\end{figure}

$(4) Example$ $4$: In this experiment, we use a real acquired signal of electrocardiogram(ECG) from the data set shared in \cite{ECGsignal}. From Fig. \ref{ECG}a in time domain, we can obviously see that this signal rides on a low power trend oscillating slowly, and also contains a large component with high frequency, sourced from the heartbeats. Observing the spectrum, significantly, there are amount of peaks, the highest one is from the trend, and then followed by a series of small components. The difficulty to decompose the real ECG signal is that excluding the peak originated from the trend, other peaks are very small, even comparable with the noise. Since without prior set-up for mode number in the new proposed algorithm, to obtain more details, there will be a large number of modes in the results, and most of them with low power.

\begin{figure}
	\centering
	\includegraphics[width=3.5in]{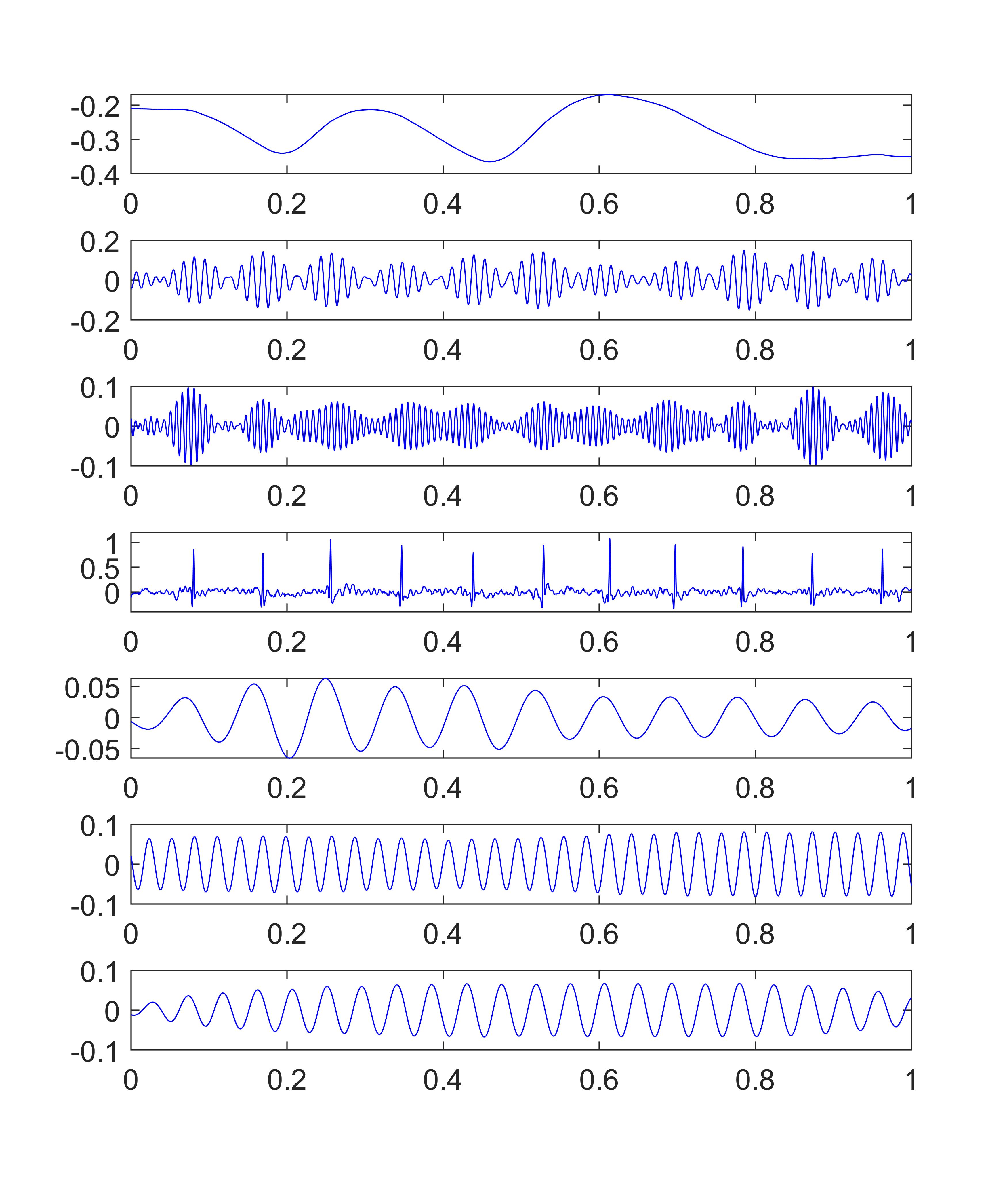}
	\caption{ECG signal decomposition}
	\label{Decomp_ECG}
\end{figure}

Unlike tests before, Fig. \ref{Decomp_ECG} shows only some selected components from the obtained mode set. During our test, there are 20 components being detected under our stop criterion, with different frequencies and amplitudes. After filtering these components with too high frequency or low amplitude and applying the refinement operation, 7 modes have been emerged and listed in Fig. \ref{Decomp_ECG}. Among them, C1 is probably the trend or baseline; and C4 is a pulse train, recording of the heartbeats; C2 and C3 can be considered as some resonances with different frequencies induced by the heartbeats, since their wave packet numbers and peak locations nearly coincide with C4; C5 oscillates like a forced vibration caused by C4, with rather low amplitude. By modifying the stop criterion, results with different mode numbers would be obtained. With rare prior knowledge, it is difficult to estimate the quality of the extracted modes, such work must be conducted by professionals. However, this test proves that our algorithm can act as a powerful tool to aid the diagnosis in the medical field.

In fact, we also conducted the decomposition for the above signals by using the EMD method. But due to the complexity of signal and noise interference, its performances are not so good in general, even can not be comparable with the other two after some manual and complicated post-processing. The most serious issue for EMD is that, there are always too many modes been extracted from the original signal, such that a single true mode is divided into pieces, and thus distributed in several extracted components with others.

\section{DISCUSSION AND CONCLUSION}
This paper has introduced a new decomposition algorithm that can extract the components of a mixture of non-stationary signals sequentially without given prior information about the modal number. This novel algorithm is based on the variational method and Hilbert transform, inspired by the VMD method. Under the fundamental assumption of narrowband, we deduced the optimization function directly from the frequency spectrum, and to achieve sequential separation purpose, a term concerning the remaining spectrum is invented and added in. These approaches are totally different from VMD. The concept of intrinsic mode functions(IMFs) has also been redefined and generalized, aiming to include typical trends.

The core of this novel algorithm is the thought of limiting the bandwidth of remaining components. It referred to the idea of dichotomy in the bandwidth consideration, but two parts with different weighting factors. Differing also from VMD, there is unnecessary to shift the signal to baseband. Naturally, we take advantage of the specific properties of Hilbert transform. Solving this optimization equation is also a simple iteration of two layers.  During a single iteration, center frequencies and recovered components update in an alternative way.

We also investigated the convergence of this algorithm. The same as VMD, our method also grantee only a local extreme, to make sure converging correctly to the expected modes, proper initialization condition is critical. By testing on a concrete signal, we find that successful initial location can be arranged in an interval near the corresponding peaks, and such interval will broaden as decomposing processes, making convergence easier to achieve. Furthermore, a proper power of noise in the background is helpful for the convergence. From the test concerning convergence interval, we have found that when the initial location beyond the corresponding correct range, the result may convergent to the neighbor modes, such that the output order of components in our algorithm also can be manipulated. 

The most important breakthrough of our algorithm is the sequential extraction for the modes under no information of mode number. In practice, when confronting a totally unknown signal, it is difficult to set the component number of the mixture, and the wrong assumption would bring mode mixing or mode dividing in VMD. Combined with the advantage of tunable output order, this algorithm is more self-adaptive than VMD in nature. Besides these advantages, utilizing some tricks, this algorithm realizes refinement and end effect reduction. Refining operation can help nearly double the performance according to our test, and the end effect reduction approach can greatly lower the deviations from the true modes in the end regions.

During our experiments, we mainly focus on non-stationary signal mixtures with noise. From the challenging tests targeting the signals of various types and with different modal numbers, we found that the so-called SVMD algorithm is superior to VMD in general, especially in reducing average deviation and suppressing end effect. However, EMD performs inferiorly to the other two, even not acceptable with some post-processing. Signals like polynomial trend and composited harmonics, violate slightly narrowband assumption and cause overlaps in the whole spectrum. However, our method also can overcome this obstacle, thus shows good robustness. We also use this algorithm to analyze a real ECG signal. Due to the sharp and strong pulse train, such signal characters with high non-stationarity. Many modes have been extracted, refinement approach clusters and merges these modes, and finally, the algorithm provides a reasonable result.

Research of this algorithm inspires us to deduce general models for the similar mode decomposition problems. Also this problem can be divided into two types: with and without mode number priori information. For VMD, a common optimization model can be expressed as:
\begin{equation}
\label{GenVMD}
L(\{\hat{u_i}\})=\left\|\hat{f}-\sum{\hat{u}_i}\right\|^2+\alpha\sum{\left\|g(\hat{u}_i,x)\right\|}^2
\end{equation}
$\hat{u}_i$ is the component, $x$ is the independent variable. $g(\hat{u}_i,x)$ is specific feature of $\hat{u}_i$, related with both $\hat{u}_i$ and $x$. $\hat{f}$ is the mixture, obtained after certain transform to the raw signal $f$, aiming to make the components more separable, like Hilbert transform and Fourier transform in this case. Of course, this operation may be unnecessary if components in the original domain are distinguishable enough. And for SVMD, such model can be rewritten as:
\begin{equation}
\label{GenSVMD}
L(\hat{u}_i)=\left\|\hat{f}_{i-1}^{r}-\hat{u}_i\right\|^2+\alpha{\left\|g(\hat{u}_i,x)\right\|}^2+\beta{\left\|g(\hat{f}_{i}^{r},x)\right\|}^2
\end{equation}
$\hat{f}_{i-1}^{r}$ and $\hat{f}_{i}^{r}$  are the $(i-1)^{th}$ and $i^{th}$ remaining signals. We can name these two models as "General Variational Mode Decomposition(GVMD)" and "Sequential General Variational Mode Decomposition(SGVMD)" methods. As a consequence, such problem has been changed to search the feature function $g$ that can describe exactly the mode character. Following this general structure, we have obtained some excellent results in similar decomposing problems, which will be discussed in later literatures.

Though with these outstanding achievements, the SVMD algorithm scarifies certain level of separating capability, which means that if the mode overlapping is strong, the decomposing results may worsen, even inferior to VMD. This issue also exists more or less in VMD. Another vital issue is the selection of penalty factors. Proper set-up for their values is critical for this multi-mode decomposition problem.

By doing some slight modifications, this algorithm can also solve the decomposition problem for 2D signal, like VMD done in \cite{VMD_2D}. Some relevant results in image processing filed have been obtained and will be presented in future papers.

\bibliographystyle{ieeetr}
 
\bibliography{ref}
\end{document}